\documentclass[aip,jcp,preprint]{revtex4-1}
\usepackage{amsmath}
\usepackage{amssymb}
\usepackage{graphicx}\graphicspath{{figures/}}
\usepackage[colorlinks=true,citecolor=blue]{hyperref}

\begin{document}

\title{Effect of surface temperature on quantum dynamics of H$_2$ on Cu(111) using a chemically accurate potential energy surface}
\author{Joy Dutta}
\altaffiliation{Contributed equally to this work}
\author{Souvik Mandal}
\altaffiliation{Contributed equally to this work}
\author{Satrajit Adhikari}
\email[Corresponding author: ]{pcsa@iacs.res.in}
\affiliation{School of Chemical Sciences, Indian Association for the Cultivation of Science, Jadavpur, Kolkata - 700 032, INDIA}
\author{Paul Spiering}
\author{J\"org Meyer}
\author{Mark F. Somers}
\affiliation{Leiden Institute of Chemistry, Gorlaeus Laboratories, Leiden University, P.O. Box 9502, 2300 RA Leiden, The Netherlands}
\begin{abstract}
The effect of surface atom vibrations for H$_2$ scattering from a Cu(111) surface at different temperatures is being investigated for hydrogen molecules in their rovibrational ground state ($v$=0, $j$=0). We assume weakly correlated interactions between molecular degrees of freedom and surface modes through a Hartree product type wavefunction. While constructing the six dimensional effective Hamiltonian, we employ: (a) a chemically accurate potential energy surface according to the Static Corrugation Model [Wijzenbroek and Somers, J. Chem. Phys. \textbf{137}, 054703 (2012)]; (b) normal mode frequencies and displacement vectors calculated with different surface atom interaction potentials within a cluster approximation; (c) initial state distributions for the vibrational modes according to Bose-Einstein probability factors. We carry out 6D quantum dynamics with the so-constructed effective Hamiltonian, and analyze sticking and state-to-state scattering probabilities. The surface atom vibrations affect the chemisorption dynamics. The results show physically meaningful trends both for reaction as well as scattering probabilities compared to experimental and other theoretical results.
\end{abstract}
\preprint{published in J. Chem. Phys. {\bf 154}, 104103 (2021), DOI:\href{http://dx.doi.org/10.1063/5.0035830}{10.1063/5.0035830}}
\maketitle

\section{Introduction}

Surface phenomena play important roles in various chemical and physical processes such as heterogeneous catalysis, growth of semiconductor devices, corrosion and hydrogen storage in metals, etc. As a result of widespread relevance, the nature and mechanism of gas phase chemical reactions on surface has been extensively studied experimentally\cite{GAA,EWG,HA2,CTH,CTD,EXPNEW1,EXPNEW2,EXPNEW3} as well as theoretically  \cite{KR4,HO1,SNB,AKT,AKT1,GD2,SC1,MBC,GJ3,HA1,ACL,SAL,SAL1,THEORYNEW1,THEORYNEW2,THEORYNEW3,SAG,TSS,TSS1,TSS2,TSSM} during past few decades. The theoretical developments on the computation of potential energy surfaces (PESs)\cite{MWMF,PSMFS, SC1} and the formulation of molecular dynamics methodologies  have  been  progressed substantially with the advancement  of  experimental  techniques,  particularly  associative  desorption  and  molecular  beam  experiments. \textit{Ab initio} molecular dynamics (AIMD) calculations employing the specific reaction parameter (SRP) approach to density functional theory for the dissociative chemisorption of D$_2$ on Cu(111) at high surface temperature (T$_\text{s}$=925K) has been performed by \citet{FNGJK1}, whereas Rettner \textit{et al.}\cite{CTR,CTR2,CTD,CTH} and Michealson \textit{et al.} \cite{HAM, HAM1, HA2} measured experimental sticking probabilities for various initial state of H$_2$/D$_2$ ($v^{\prime}$, $j^{\prime}$) - Cu(111) systems. 
Recently, Wodtke and coworkers\cite{Wodtke} experimentally observed an unusual slow channel along with the mostly common fast one for the dissociative adsorption of H$_2$/D$_2$ on Cu(111)/Cu(211) around low kinetic energies (below 0.2 eV) of the incoming diatom at higher surface temperature (T$_\text{s}$=923$\pm$3 K). Such unusual channel indicates an interesting additional reaction mechanism, where trapped reactant tunnels through a substantial barrier much before attaining the vibrational equilibrium state (thermal equilibrium) due to the involvement of thermal fluctuation of Cu(111)/Cu(211) surface.  

Construction of accurate PESs has been a topic of interest in the regime of molecule-surface scattering processes. Wiesenekker, Kroes, and Baerends\cite{GGAGJK} developed a six dimensional (6D) PES using the generalized gradient approximation (GGA) of density functional theory (DFT) for describing dissociative chemisorption of H$_2$ over Cu(100) surface. On the other hand, a more chemically accurate 6D PES was constructed by D\'{i}az {\it et al.}\cite{SC1,CDGJK} employing SRP\cite{SRPKROES} approach on DFT. Quantum(Q)/quasi-classical(QC) dynamical calculations had been performed under Born-Oppenheimer static surface (BOSS) approximation to investigate state-resolved dissociative chemisorption probabilities as a function of collisional energy for  H$_2 /$D$_2$ ($v$, $j$)-Cu(111) systems.

The effect of surface temperature on reaction probability in gas-metal surface collision processes is one of the most fascinating phenomena, which has been explored with different theoretical approaches. 
For example, AIMD relies on QC trajectories to take into account the surface temperature effect, where the motion of surface atoms are simulated through ``on the fly" calculation of  forces. 
In particular, Nattino \textit{et al.}\cite{FNGJK1} have shown that the use of sufficiently flexible  asymmetric sigmoidal generalized logistic function (LGS) for fitting the raw time-of-flight TOF spectra provides more accurately fitted experimental reaction probability curves with different saturation values at high collisional energies. At 925K, AIMD calculations demonstrate that theoretical dissociation probability profiles for D$_2 \: (v, j)$-Cu(111) systems are close to experimental observations only at low collision energies, but at high collision energy range, theoretical results are higher in magnitude than the experimental ones. Moreover, broadening of reaction probability with AIMD is much smaller compared to experimental data\cite{HAM}. 

On the other hand, Wijzenbroek and Somers\cite{MWMF} constructed a static corrugation model (SCM) for dissociation of H$_2$/D$_2$($v$, $j$) on Cu(111). The SCM incorporates surface temperature effects by considering thermal expansion and thermal displacements of surface atoms \cite{AMGJK} within a vibrational sudden approximation for the dynamics, which are then carried out based on an effectively six-dimensional PES. The resulting QC dynamics has been compared with BOSS, AIMD methods and experimental data\cite{MWMF}. Furthermore, Spiering, Wijzenbroek and Somers\cite{PSMFS} extended the original SCM model by including effective three-body interactions, a corrected surface stretching scheme, and fitting the model to additional DFT data for chemisorption of D$_2$ on Cu(111). 

In the last few years, construction of chemically accurate high-dimensional neural network potentials (HD-NNPs)\cite{PIP-NN,PIP-NN1,PIP-NN2,AtNNmethod,AtNN,AtNN1} for various important gas-metal collisional processes (e.g., CO$_2$-Ni(100)/Pt(111)\cite{CO2-Ni,CO2-Ni-new,CO2-Pt}, NO-Au(111)\cite{NO-Au}) has been progressed extensively to overcome the bottleneck of expensive AIMD method. Such neural network based approach allows accurate calculation of reaction probabilities even with very low magnitude ($10^{-5}-10^{-4}$) for highly activated chemisorption reactions, N$_2$ + Ru(0001)\cite{N2-Ru} and CHD$_3$ + Cu(111)\cite{CHD3-Cu}. Recently, Jiang \textit{et al.}\cite{H2-Cu-multiple} reported a universal highly transferable PES by employing a newly developed embedded atom neural network (EANN)\cite{EANN} approach for dissociative chemisorption of H$_2$ on multiple low-index copper surfaces [Cu(111)/Cu(100)/Cu(110)]. The novel EANN PES allows to determine quantitative surface temperature (T$_s$) dependent barrier distributions and thereby, enables to explore crucial role of thermal expansion effect.
However, currently to the best of our knowledge, although 6D QD reaction probabilities were estimated by employing a direct reactive flux method on HD-NNPs\cite{H2-Cu-multiple,6D-QD-HD-NNP-B-Jiang-H-Guo,6D-QD-HD-NNP-B-Jiang-H-Guo-1,6D-QD-HD-NNP-D-Zhang,6D-QD-HD-NNP-D-Zhang-1,6D-QD-HD-NNP-D-Zhang-2} for various systems, QD calculations have not been attempted so far on HD-NNPs to obtain converged inelastic scattering and diffraction probabilities. It remains to be seen if currently available implementations of HD-NN codes are fast enough to be able to do this with the same accuracy as traditional corrugation reducing procedure\cite{CRP} (CRP) PESs\cite{CRP1,AMGJK,CRP2,CRP3,MWMF,PSMFS,CRP4} have shown to offer in numerous cases.

Although several first-principles-based theoretical attempts have been made to unveil the effect of surface temperature and its connections to surface vibrations and electronic excitations on molecule-surface scattering processes, the theoretical outcomes are still far away from the actual experimental observations. The following types of broad theoretical approaches have been implemented in the dynamical calculations including the surface mode(s) to account for surface temperature effects: (a) A single or few surface oscillator (SO) \cite{ACL,HA1,SAL,HO1,SAL1,MDPS,UZER1,MILLER,SALIN} models have been adapted to construct the Hamiltonians for H$_2$-Cu(1nn)/Si(100) systems. Also, theoretical approaches have been improved by considering modified surface oscillator models (MSO)\cite{SAL,UZER1}; (b) Nave and Jackson investigated\cite{SNB,JACKSON1,AKT1,AKT} the role of lattice motion\cite{MBC} and reconstruction for CH$_4$ dissociation on Ni(111) plane on a 4D PES at various temperatures within the harmonic approximation; (c) Adhikari and co-workers\cite{SAG,TSS,TSS1,TSS2,TSSM} have carried out 4D$\otimes$2D and 6D QD for H$_2$/D$_2$ ($v$, $j$)-Cu(1nn)/Ni(100) systems by employing TDDVR-methodology\cite{TSS1,TSSM,SM-MOLPHYS,SMIRPC} on more realistic many oscillator\cite{GD2,SAG,TSS,TSS1,TSS2,TSSM} model mimicking a specific plane (1nn) of a metal surface [Cu(1nn)/Ni(100)]. In those approaches, the effective Hamiltonian has been formulated under the mean-field\cite{TSS,TSS1,TSS2,TSSM,SM-JTCC,SM-MOLPHYS} approximation assuming weak coupling among molecular degrees of freedom (DOFs) and surface modes. The vibrational frequencies are computed from a metal-metal EDIM-fitted \cite{TNT} potential, while their distributions at the specific temperature are incorporated through the Bose-Einstein (BE) or Maxwell-Boltzmann (MB) probability factor. Although the reaction probabilities obtained from this 6D QD calculations could show up broadening\cite{SAG,TSS,TSS1,TSS2,TSSM} as compared to experimental observations at higher surface temperature (T$_\text{s}$=1000K), the sigmoid nature and the appropriate inflection point of the experimental fitted curves\cite{CTR} are absent in the theoretical ones.  
 
In this present work, the surface temperature effect on the transition/reaction probability of H$_2$($v$=0, $j$=0)-Cu(111) system has been investigated more critically by combining a first principle based many oscillator model\cite{GD2,SAG,TSS,TSS1,TSS2,TSSM} with a chemically accurate PES from the SCM\cite{MWMF} relying on a mean-field approximation. We have reformulated an effective Hamiltonian by considering the solutions of linearly forced harmonic oscillator (LFHO)\cite{PPJS}, where the surface temperature has been incorporated by taking into account the BE or MB probability factors for the initial state distribution of those modes. The surface mode frequency spectrum and displacement vectors are modeled by a cluster approximation, where the interaction between the copper atoms is described by different potentials. The interaction potential between molecular (H$_2$) DOFs and surface (Cu(111)) modes as well as its first derivatives are obtained from the SCM potential. The scattering calculations (6D) have been carried out with split operator (SPO)-DVR QD code\cite{SPO} to obtain transition as well as reaction probability of H$_2$ ($v$=0, $j$=0) on Cu(111) surface. Finally, we show reaction and vibrational-state-resolved scattering probabilities in comparison with other theoretical and available experimental results.

\section{Theoretical Background}

An effective Hamiltonian has been formulated by invoking a mean-field approach to incorporate the effect of surface vibrational modes at non-zero surface temperature by introducing the BE probability factors for their initial state distribution. Such an approach allows time evolution of molecular degrees of freedom (DOFs) ($\{X_k\}$) as well as surface modes ($\lbrace Q\rbrace$) to access all possible configurations arising from their various quantum states, where each subsystem ($\{X_k\}/\lbrace Q\rbrace$) is fully correlated with its all possible configurations. Due to the huge mass difference between the diatom and the metal atoms, the cross correlations among the configurations of different subsystems are neglected by assuming weak interaction and thereby, a product type wavefunction is considered as given bellow:
\begin{eqnarray}
\Psi(x, y, z, X, Y, Z, t)\cdot \Phi_\text{vib}(\{Q\},t),
\label{eq:meanfield}
\end{eqnarray}

\begin{figure}
\centering
\includegraphics[width=\columnwidth]{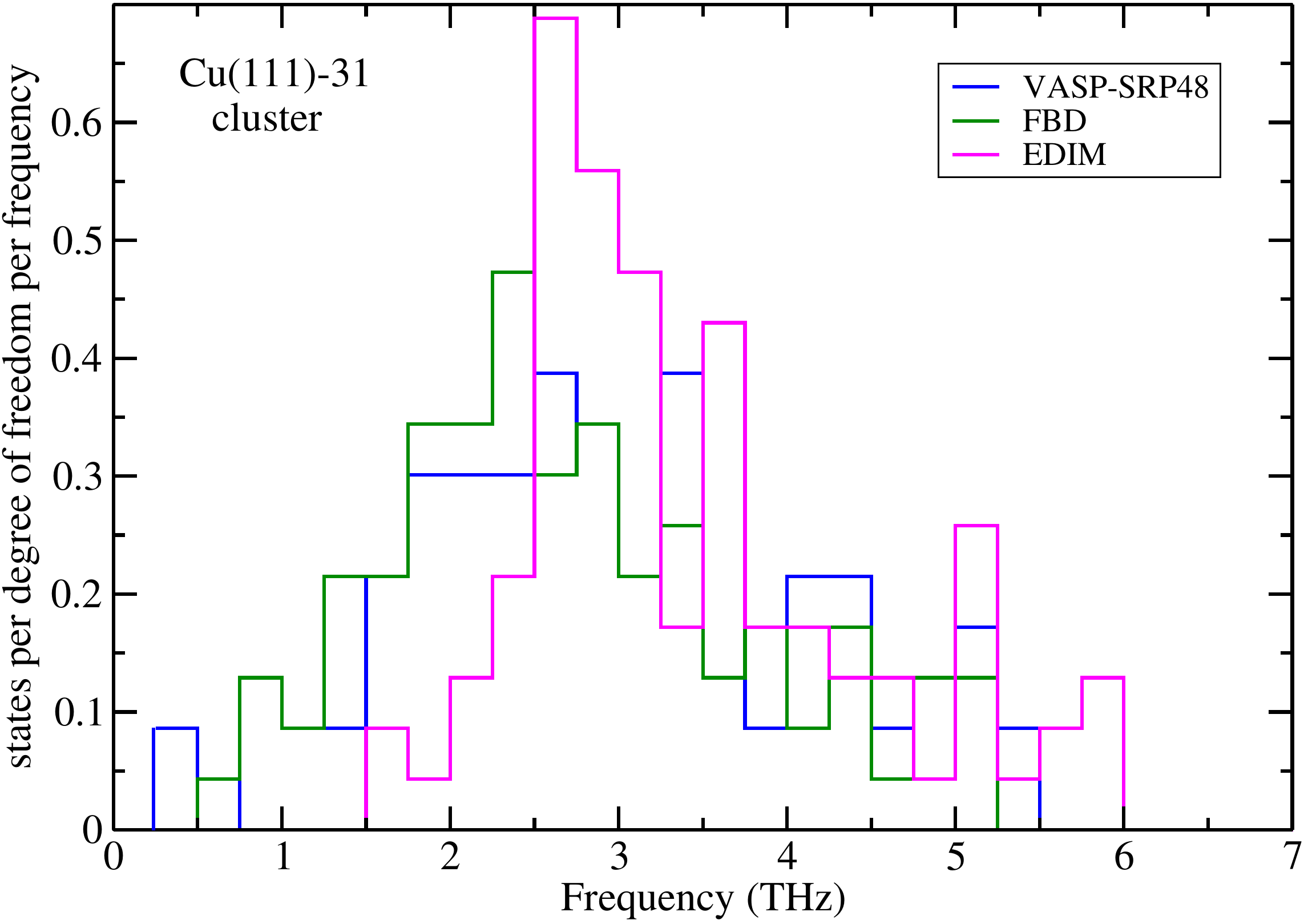}
\caption{Frequency distributions of the Cu$_{31}$ cluster models as described in the text, calculated with the VASP-SRP48, FBD and EDIM potentials.}
\label{fig:histogram}
\end{figure}

The diatomic molecule has six DOFs denoted by $(x, y, z, X, Y, Z)$. Here the Cartesian coordinates $x$, $y$, $z$ represent the molecular vector $\mathbf{R} = (R, \theta, \phi)$ and $X$, $Y$, $Z$ are the center of mass of the diatom with respect to the Cu surface such that the top layer of the Cu atoms corresponds to $Z$ = 0.
Surface vibrational wavefunctions ($\Phi_\text{vib}(\{Q\})$) and their concomitant frequencies ($\lbrace \omega_k \rbrace$) are modeled by the $87$ ($=3*31 - 6$) normal modes of a Cu$_{31}$ cluster that has been cut out of the topmost three layer of the Cu(111) surface. The interaction between the copper atoms is described by the SRP48 DFT functional as implemented in VASP\cite{SC1} (VASP-SRP48), the Embedded-atom method (EAM) potential originally developed by Folies, Baskes and Daw (FBD)\cite{FBD} 
and a potential based on the Embedded Diatomics in Molecules (EDIM) model\cite{TNT}.
Further computational details about these frequency calculations are described in the supporting information. 
Figure \ref{fig:histogram} shows the three different resulting frequency spectra.

The product-type wavefunction [Eq. (\ref{eq:meanfield})] leads to the following form of time and temperature dependent effective Hamiltonian:
\begin{eqnarray}
\widehat{H} (x, y, z, X, Y, Z, t, T_\text{s}) =&-&\frac{\hbar^2}{2\mu} \bigg(\frac{\partial^2}{\partial x^2}
+ \frac{\partial^2}{\partial y^2} + \frac{\partial^2}{\partial z^2}\bigg)
-\frac{\hbar^2}{2M} \bigg(\frac{\partial^2}{\partial X^2}
+ \frac{\partial^2}{\partial Y^2} + \frac{\partial^2}{\partial Z^2}\bigg)\nonumber\\
&+& V_0(x, y, z, X, Y, Z) + V_\text{eff}(x, y, z, X, Y, Z, t, T_\text{s}),
\label{eq:hamiltonian}
\end{eqnarray}
where $\mu$ and $M$ are the reduced and total mass of the diatom, respectively. $V_0(x, y, z, X, Y, Z)$ is the rigid surface (RS)-molecule interaction potential known as the BOSS PES\cite{SC1,CDGJK} describing the situation of ideal static lattice (i.e., excluding the effect of surface DOFs), where $V_\text{eff}(x, y, z, X, Y, Z, t, T_\text{s})$ is the effective Hartree potential due to the surface mode coupling with molecular DOFs.

It is worth mentioning that although the dynamics of the molecule is characterized by a single wavepacket evolution (apparently pure state representation), the effective Hartree potential $V_\text{eff}$ arising from molecular DOFs and surface modes (bath) coupling is constructed by taking into account an ensemble average of different pure state configurations (i.e., mixed state situation) through the employment of the MB/BE probability factor and consequently, the surface temperature is introduced into the effective Hamiltonian parametrically. Therefore, such product type of Hartree wavefunction description could simulate the molecular beam experimental situation, where molecule and surface initially are not in thermal equilibrium with each other.

\subsection{Formulation of the effective Hartree potential ($V_\text{eff}$)}

The effective Hartree potential averaged over initial state $\lbrace n_0 \rbrace$ distribution of vibrational modes is defined as (see Appendix A)
\begin{eqnarray}
V_\text{eff}(x, y, z, X, Y, Z, t, T_\text{s})=\langle V \rangle (t, T_\text{s}) = \sum_{\lbrace n_0 \rbrace} p_{\lbrace n_0 \rbrace} \langle V \rangle_{\lbrace n_0 \rbrace},
\label{Eq: state_averaged_Hartree_potential}
\end{eqnarray}
where the initial states are averaged out by considering the Bose-Einstein (BE) or Maxwell-Boltzmann (MB) distribution over various surface mode ($k$)
\begin{eqnarray}
p_{\lbrace n_0 \rbrace} = \prod_{k=1}^M p_{n_k^0}^{(k)}.
\end{eqnarray}
The initial state $\lbrace n_0 \rbrace$ dependent Hartree potential due to the interaction potential ($V_\text{I}$) between gas molecular DOFs and surface modes is expressed as: 
\begin{eqnarray}
\langle V \rangle_{\lbrace n_0 \rbrace} &=& \langle \Psi(t) \vert V_\text{I} \vert \Psi(t) \rangle	\nonumber  \\
&=& \sum_{\lbrace n^{\prime} \rbrace} \sum_{\lbrace n \rbrace} \alpha^{*}_{\lbrace n^{\prime}\rbrace \leftarrow \lbrace 
n_{0}\rbrace} (t) \:
\alpha_{\lbrace n \rbrace \leftarrow \lbrace n_{0}\rbrace} (t) \langle \lbrace n^{\prime} \rbrace \vert V_\text{I} \vert \lbrace n 
\rbrace \rangle,
\label{Eq: Initial_V_eff}
\end{eqnarray}
where the $\lbrace n\rbrace$ and $\lbrace n^{\prime}\rbrace$ are the all possible quantum states accessible by the vibrational modes. 
The amplitudes $\alpha_{\lbrace n \rbrace} (t)$ originating from initial state ${\lbrace n_0 \rbrace}$ to final ones ${\lbrace n \rbrace}$ under the evolution operator, $U(t, t_0)$, due to molecule-surface interaction is defined as:
\begin{eqnarray}
\alpha_{\lbrace n \rbrace \leftarrow \lbrace n_{0}\rbrace} (t) = \langle \lbrace n  \rbrace \vert U \vert \lbrace n_0  \rbrace 
\rangle.
\label{Eq: amplitude_state}
\end{eqnarray}
While averaging over all possible final quantum states ($\lbrace n\rbrace$ and $\lbrace n^{\prime}\rbrace$) in Eq. (\ref{Eq: Initial_V_eff}), the interaction potential ($V_\text{I}$) between  molecular DOFs and surface modes needs to be expanded in terms of normal mode coordinate ($Q_k$). Furthermore, those coordinates are expressed in terms of boson creation ($b_k^{\dagger}$)/annihilation ($b_k$) operators such as $Q_k = A_k (b_k^{\dagger} + b_k)$ and $A_k = \sqrt{\hbar/{2\omega_k}}$ and thereby, the interaction potential considering only up to first order terms, takes on the following form: 
\begin{eqnarray}
V_\text{I}=V_{0} + \sum^M_{k=1}  \lambda_k A_k(b_k F^-_k + b^+_k F^+_k) V_{k,1}, 
\label{Eq: Interaction_potential_Interaction_picture}
\end{eqnarray}
where $F^-_k = \exp(-i\omega_k t)$, $F_k^+= (F_k^-)^\ast$  
and $V_{k,1} = \partial V_\text{I}/\partial Q_k|_\text{eq}$.
Finally, the first derivative of interaction potential ($V_{k,1}$) with respect to normal mode 
($Q_k$) is evaluated by employing the chain rule of differentiation w.r.t. metal atomic position 
$\left(\dfrac{\partial V_{a\alpha}^\text{Cu-H}}{\partial X_{\alpha i}}\right)$ as given below:
\begin{small}
\begin{eqnarray}
V_{k,1}=\left(\frac{\partial V_{I}}{\partial Q_k}\right)
&=&\sum_{a\alpha}\dfrac{\partial(V_{a\alpha}^\text{Cu-H}(r_{a\alpha})-V_{a\alpha}^\text{Cu-H}(r_{a\alpha}^\text{id}))}{\partial 
Q_{k}} \nonumber\\
&=&\sum_{a\alpha i}\left[\dfrac{\partial V_{a\alpha}^\text{Cu-H}(r_{a\alpha})}{\partial X_{\alpha i}}-\dfrac{\partial 
V_{a\alpha}^\text{Cu-H}(r_{a\alpha}^\text{id})}{\partial X_{\alpha i}}\right].
\dfrac{\partial X_{\alpha i}}{\partial Q_{k}}\nonumber \\
&=&\sum_{a\alpha i}m_{\alpha}^{-1/2}\left[\dfrac{\partial V_{a\alpha}^\text{Cu-H}(r_{a\alpha})}{\partial X_{\alpha i}}-\dfrac{\partial V_{a\alpha}^\text{Cu-H}(r_{a\alpha}^\text{id})}{\partial X_{\alpha i}}\right]T_{\alpha i;k}
\label{potfirst}
\end{eqnarray}
\end{small}
where the following equation is used to calculate $\dfrac{\partial X_{\alpha i}}{\partial Q_{k}}$:
\begin{eqnarray}
X_{\alpha i}-X_{\alpha i}^\text{id}&=&m_{\alpha}^{-1/2}\sum_{k}T_{\alpha i;k}Q_{k}.
\label{transformation}
\end{eqnarray} 

and the derivative of the SCM potential\cite{MWMF} $\left(\dfrac{\partial V_{a\alpha}^\text{Cu-H}}{\partial X_{\alpha i}}\right)$ is shown in Appendix B. 
The indices $\alpha$, $a$ and $k$ denote metal atom, gas atom and normal mode, respectively, where $\alpha$ = 1, 2 $\cdots$ $N$, $N$ = 31 (no. of metal atoms), $a$ = 1, 2 and $k$ =  7, 8 $\cdots$ $3N$, where the first 6 modes are the translational and rotational DOFs. $X_{\alpha i}$ is the position of a metal atom, where $X_{\alpha i}^\text{id}$ is its equilibrium position for a specific degree of freedom, $i$. $Q_{k}$ is the normal mode coordinate, $m_{\alpha}$ is the mass of surface atom and $T_{\alpha i; k}$ is the transformation matrix between local ($\alpha i$) and normal modes ($k$). On the other hand, $r_{a\alpha}$ is the distance between each metal ($\alpha$) and gas ($a$) atom. $V_{a\alpha}^\text{Cu-H} (r_{a\alpha})$ and $V_{a\alpha}^\text{Cu-H}(r_{a\alpha}^\text{id})$ are the gas-metal interaction potentials due to displaced and ideal positions of the metal atoms, respectively. 
Inserting Eq. (\ref{Eq: Interaction_potential_Interaction_picture}) in Eq. (\ref{Eq: Initial_V_eff}), and then in Eq. (\ref{Eq: state_averaged_Hartree_potential}) for the BE or MB cases,  we arrive at the following compact form of the effective Hartree potential: 
\begin{small}
\begin{eqnarray}
V_\text{eff}^\text{BE} (x,y,z, X, Y, Z, t, T_\text{s}) = \frac{1}{N_\text{BE}} \left[\sum_{k=7}^{3N} \lambda_k \frac{1}{\omega_k^2} 
V_{k,1}^2 \left[ \cos \omega_k (t - t_0) - 1 \right]\
 \sum^{\infty}_{q=1}\frac{ z_k^{q/2}}{(1-z^q_k)} \right],
\label{Eq:V_Hartree_final_form_I1}
\end{eqnarray}
\end{small}
where $N_\text{BE} =\sum_k \sum_{n_k^0 =0}^{\infty} \frac{1}{\exp \left[ \hbar \omega_k \left( n_k^0 + \frac{1}{2} \right) \beta \right] - 1 }$ is the normalization of Bose-Einstein probability factor for vibrational modes. On the contrary, in case of the Maxwell-Boltzmann probability factor, the form of effective Hartree potential will be:
\begin{small}
\begin{eqnarray}
V_\text{eff}^\text{MB} (x,y,z, X, Y, Z, t, T_\text{s}) = \frac{1}{N_\text{MB}} \left[\sum_{k=7}^{3N} \lambda_k \frac{1}{\omega_k^2} 
V_{k,1}^2 \left[ \cos \omega_k (t - t_0) - 1 \right]\
 \frac{ z_k^{1/2}}{(1-z_k)} \right],
\label{Eq:V_Hartree_final_form_I1_MBP}
\end{eqnarray}
\end{small}
with $N_\text{MB} = \sum_k \sum_{n_k^0 =0}^{\infty}  \exp \left[ - \hbar \omega_k \left( n_k^0 + \frac{1}{2} \right) \beta\right]$. The associated sign (- or +) of the first derivative of the interaction potential is denoted by $\lambda_k$.

Some important aspects of the effective Hamiltonian: (a) Both for the BE and MB cases, the frequency
($\omega_k$) of the surface modes appear multiple times in the time and temperature dependent terms of the effective potential. Therefore, the frequency spectrum (see Figure \ref{fig:histogram}) calculated by different approaches (VASP-SRP48, FBD and EDIM) from the surface atom interaction potential is expected to play a crucial role in the reaction and scattering probabilities; (b) The functional form of temperature dependent term of the effective potential in terms of the partition function differs for the BE ($\sum^{\infty}_{q=1}\frac{ z_k^{q/2}}{(1-z^q_k)}$) and MB ($\frac{ z_k^{1/2}}{(1-z_k)}$) cases and thereby, their contributions would be different on the broadening of the reaction probabilities at a particular temperature; (c) The magnitude and the occurrence of the first derivative of the interaction potential ($V_{k,1}$) should have a role on the reaction probability (see Figure \ref{fig:3dhis}); (d) For the specific surface mode frequency ($\omega_k$) and temperature ($T_\text{s}$), the contribution of the Hartree potential is modulated as a function of time of the collision process. 

\subsection{Mean-field approach and Sudden approximation}

The theoretical description of molecular DOFs-surface mode interaction and the dynamical outcomes of gas-metal surface scattering process are described using  mean-field and sudden approximation as below:

(a) The explicit correlations between molecular DOFs and surface mode vibrations are neglected in both the approaches either by sampling the lattice vibration ($Q$) using MB distribution (sudden approximation\cite{SNB,JACKSON1,AKT1,AKT,MBC}) or by employing a Hartree product type of wavefunction (mean-field approach\cite{TSS,TSS1,TSS2,TSSM,SM-JTCC,SM-MOLPHYS}) through the construction of effective potential. In both cases, only the effective contribution of the surface mode vibrations at the particular surface temperature is taken into account on the motion of incoming molecule; (b) In the  mean-field treatment, the effective potential has been constructed by including all possible initial state configuration ($\lbrace n_0 \rbrace$) through the employment of MB/BE distribution by considering all the vibrational states for each surface mode ($k$). Such effective potential changes with time during the course of collisional process due to surface mode excitations at the particular temperature. Moreover, time dependence of the effective potential also varies for different surface temperatures and kinetic energies (KEs) of the diatom. Therefore, instantaneous effects of molecule-lattice atom interactions are incorporated in the effective potential implicitly within mean-field approach. On the other hand, such responses of the lattice atoms (e.g., instant puckering) had been considered into sudden approximation treatment by performing scattering calculations on different sampled (classically) points ($\lbrace Q \rbrace$) of the lattice vibration ($Q$) at the given surface temperature; (c) Ensemble average of scattering probabilities obtained from the sudden approximation over infinitely distinct sampled values of a large configuration space and the scattering profile resulting from the mean-field approach by taking the time average over infinitely different effective potential arising from all possible configuration, could have comparable level of approximations due to the neglect of higher order correlations between molecular DOFs and surface mode. The applicability of both approaches could be validated only through the implementation on specific system(s).
 
\subsection{Computational details for the effective Hatree potential}

The expression of effective Hartree potential contains the derivatives of the interaction potential ($V_{k,1}=\frac{\partial V_\text{I}}{\partial Q_k}$), frequency ($ \omega_k$) of surface modes ($Q_k$), time ($\left[ \cos \omega_k (t - t_0) - 1 \right]$) and the temperature ($\sum^{\infty}_{q=1}\frac{ z_k^{q/2}}{(1-z^q_k)}$ for BE, $\frac{z_k^{1/2}}{(1-z_k)}$ for MB) dependent terms (see Eq. (\ref{Eq:V_Hartree_final_form_I1}) and (\ref{Eq:V_Hartree_final_form_I1_MBP})). The derivatives of the interaction potential with respect to the normal modes ($V_{k,1}=\frac{\partial V_\text{I}}{\partial Q_k}$) have been computed with the chemically accurate SCM \cite{MWMF} potential, where the transformation matrix ($T_{\alpha i; k}$) between local ($X_{\alpha i}$) and normal modes ($Q_k$) is being employed. The frequency spectrum ($\lbrace \omega_k \rbrace$) and displacement vector ($T_{\alpha i; k}$) of surface modes ($Q_k$) have been evaluated by using the VASP-SRP48, FBD or EDIM surface atom interaction potentials. With such an effective Hamiltonian, we perform 6D QD for H$_2$ on Cu(111) start with the hydrogen molecule in its rovibrational ground state ($v=0, j=0$) using the SPO-DVR code\cite{SPO}. The parameters of the SRO-DVR code are given in section 2 of the supplementary material. We calculate reaction and scattering probabilities for various surface temperatures ($T_\text{s} =$ 1 K, 120 K, 300 K, 600 K and 925 K).

\section{Results}

\begin{figure}
\centering
\includegraphics[width=\columnwidth]{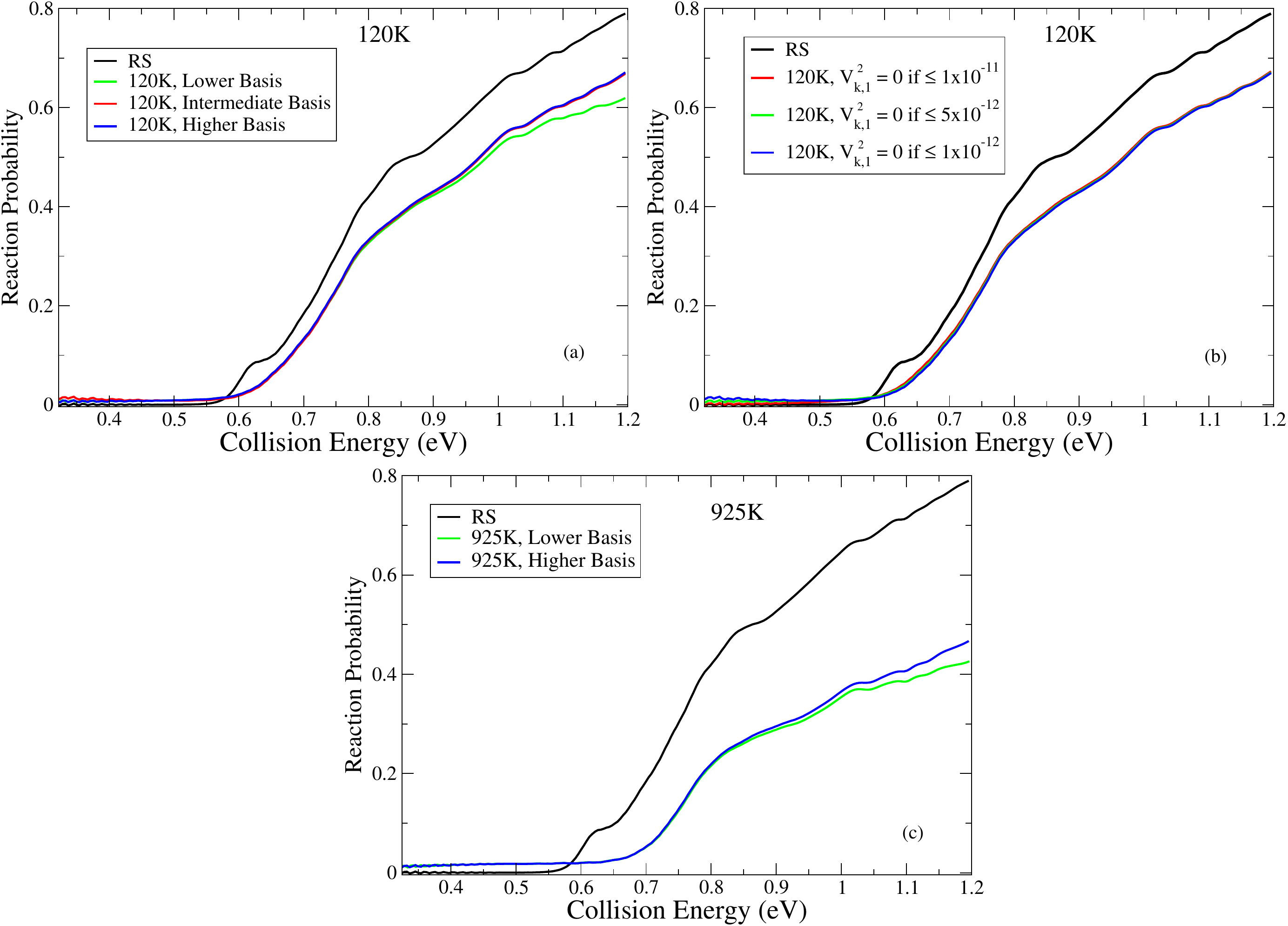}
  \caption{ Convergence of the reaction probability for H$_2$ on Cu(111) with the VASP-SRP48 calculated normal mode frequencies as a function of (a) basis set with lower ($X$ = 18,  $Y$ = 18, $Z$ = 140, $R$ = 64, $j_{\text{max}}$ = 12 and $m_{j\text{max}}$ = 6), intermediate ($X$ = 18, $Y$ = 18, $Z$ = 180, $R$ = 64, $j_{\text{max}}$ =  20 and $m_{j\text{max}}$ = 10) and higher ($X$ = 18, $Y$ = 18, $Z$ = 180, $R$ = 64, $j_{\text{max}}$ =  24 and $m_{j\text{max}}$ = 12) bases for 120 K; (b) cut-off by setting $V_{k,1}^2$ equal to zero (0) if its magnitude is $\leq$ $1.0\times 10^{-11}$,  $\leq$ $5.0\times 10^{-12}$ and $\leq$ $1.0\times 10^{-12}$, respectively at 120 K; (c) basis set with lower ($X$ = 18,  $Y$ = 18, $Z$ = 140, $R$ = 64, $j_{\text{max}}$ = 12 and $m_{j\text{max}}$ = 6) and higher ($X$ = 18, $Y$ = 18, $Z$ = 180, $R$ = 64, $j_{\text{max}}$ =  24 and $m_{j\text{max}}$ = 12) bases for 925 K surface temperature by imposing the cut-off condition, $0=V_{k,1}^2\leq1.0\times 10^{-11}$.} 
\label{fig:convergence}
\end{figure}

For the surface temperature of 120 K, using VASP-SRP48 calculated normal mode frequencies, the convergence profiles of reaction probability as a function of the basis set as well as the cut-off on the derivative of the interaction potential ($V_{k,1}^2$) are demonstrated in Figure \ref{fig:convergence}(a) and \ref{fig:convergence}(b), respectively. On the other hand, for 925 K situation, convergence test of reaction probability profiles is performed with the same basis set functions (lower and higher basis) as used in 120 K case (see Figure \ref{fig:convergence}(c)). It is worth mentioning that performing QD calculations with further larger basis set is computationally very expensive. There are two important things to note from these figures: (a) The dependence of reaction probability with lower, intermediate and higher basis sets for 120 K and with lower and higher ones for 925 K are minimum except at higher collisional energy for lower basis; (b) The different values ($1.0\times 10^{-11}$, $5.0\times 10^{-12}$ and  $1.0\times 10^{-12}$) of cut-off on the derivative of the interaction potential ($V_{k,1}^2$) does not show any effect for the case, 120 K. Moreover, it has been observed that if the magnitude of the cut-off does not show any effect within the cut-off on $V_{k,1}^2$ $\leq$ $1\times10^{-12}$, there is no effect on further lowering of the cut-off, which has been numerically verified. Whereas at 925 K surface temperature, cut-off on $V_{k,1}^2$ is imposed for the condition, $V_{k,1}^2 \leq 1.0\times 10^{-11}$. 

\subsection{Effect of quantum vs. classical initial vibrational state populations on the reaction probability}

\begin{figure}
\centering
\includegraphics[width=\columnwidth]{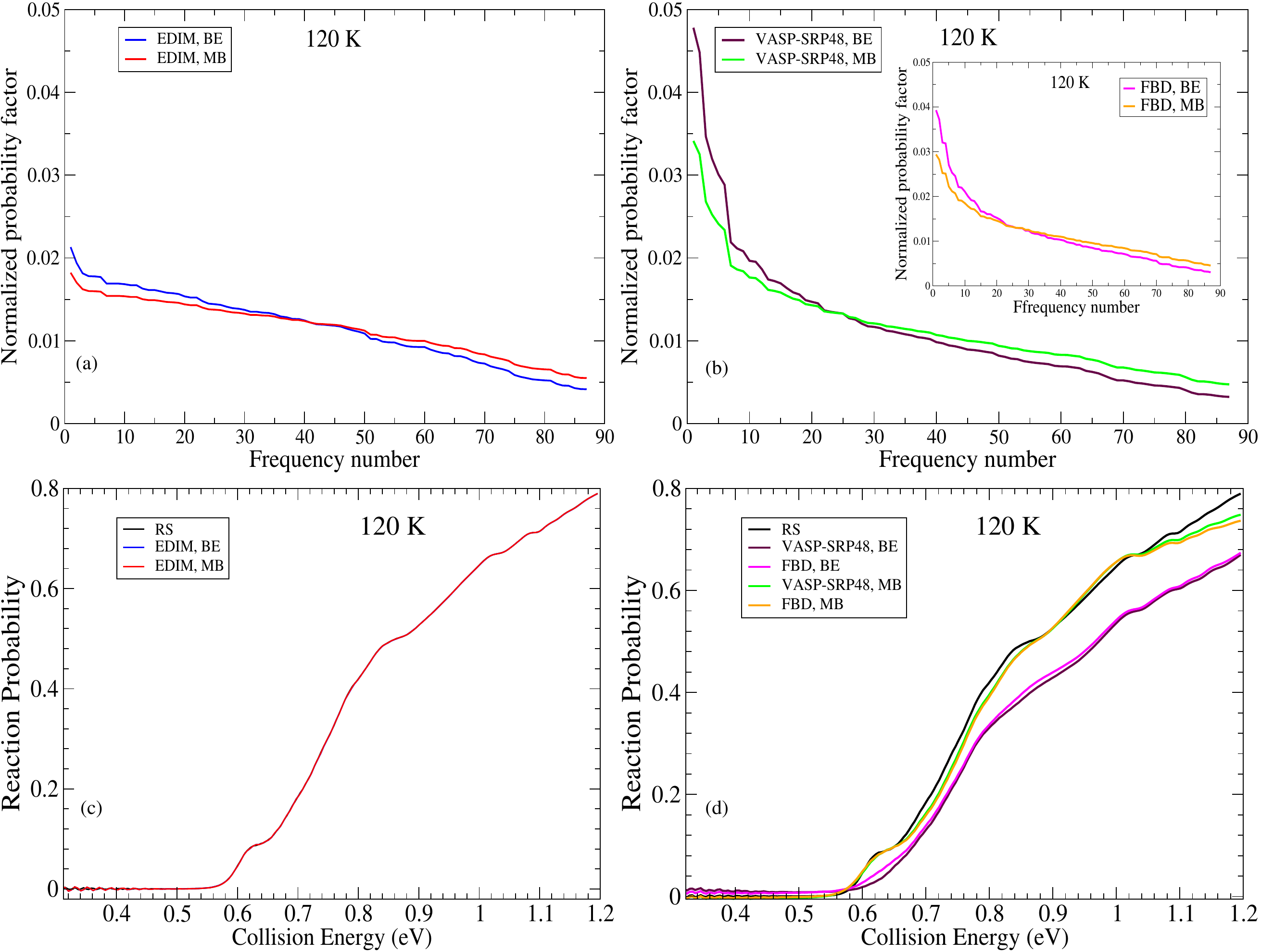}
  \caption{Normalized probability factor as a function of the frequency number with the BE and the MB distribution at 120 K for (a) the EDIM normal mode frequencies and (b) the VASP-SRP48/FBD normal mode frequencies. Reaction probabilities for H$_2$ on Cu(111) calculated based on the Hartree potential constructed with (c) the EDIM and (d) the VASP-SRP48/FBD calculated normal mode frequencies along with the MB and the BE probability factor at 120 K surface temperature.}
\label{fig:BEMB}
\end{figure}

\begin{figure}
\centering
\includegraphics[width=\columnwidth]{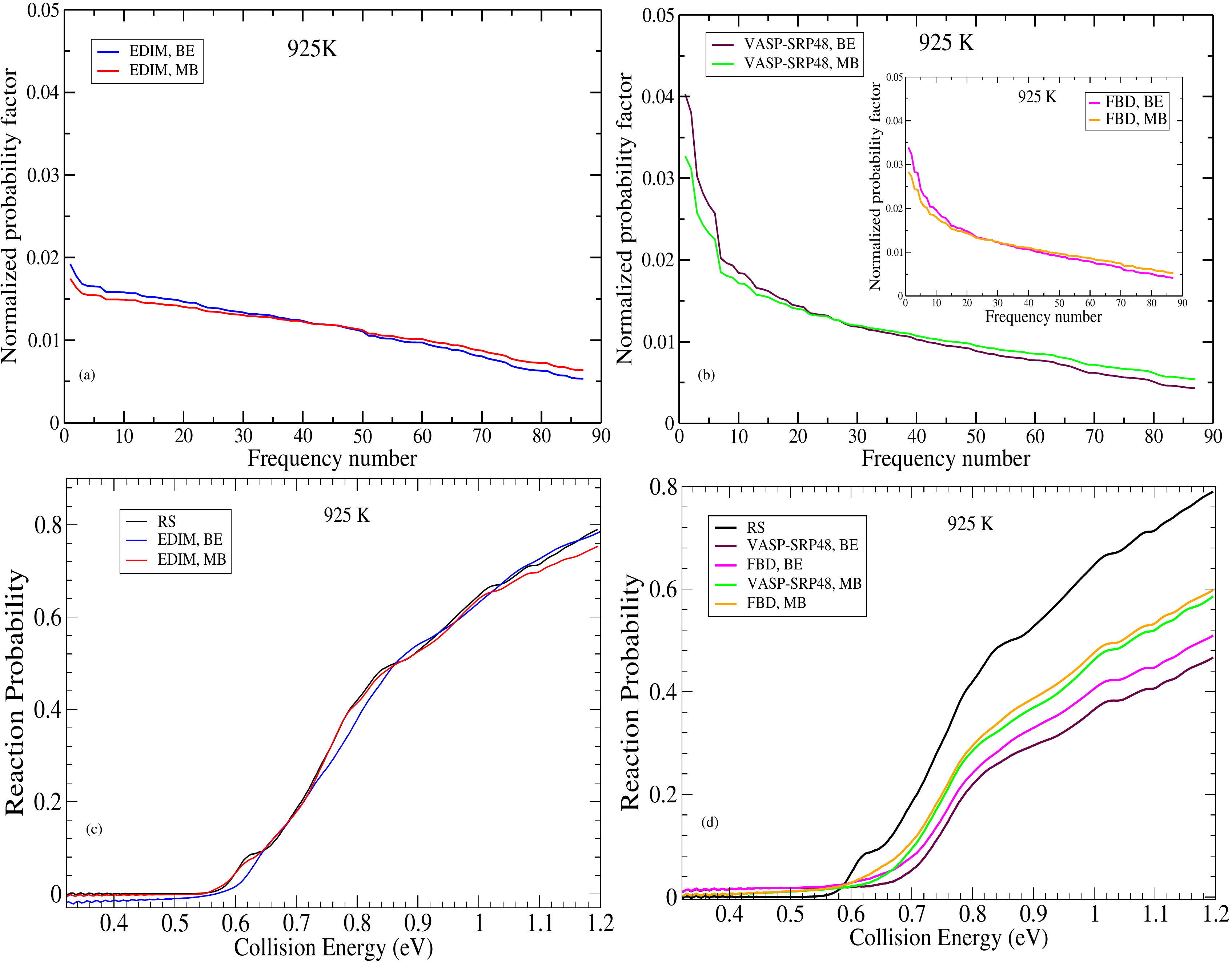}
  \caption{Same as Figure \ref{fig:BEMB}, but for the surface temperature of 925 K.}
\label{fig:BEMB925}
\end{figure}

In Figures \ref{fig:BEMB}(a)-\ref{fig:BEMB}(b) and \ref{fig:BEMB925}(a)-\ref{fig:BEMB925}(b), we have presented the contribution of normalized probability factor (see Eq. (\ref{Eq:norm_prob}) in Appendix A)  due to the BE and MB statistics as a function of frequency number calculated from the EDIM and VASP-SRP48/FBD potentials at 120 K and 925 K surface temperature, respectively. In case of the EDIM normal modes, Figure \ref{fig:BEMB}(a) and \ref{fig:BEMB925}(a) depict the magnitudes of normalized probability factor over the entire range of vibrational frequencies at 120 K and 925 K, where their values are quite low and close to each other both for the MB and BE statistics. As a result, the EDIM frequencies do not show any broadening (see Figure \ref{fig:BEMB}(c)) or have an almost negligible effect on the reaction probabilities (see Figure \ref{fig:BEMB925}(c)) either with the BE or the MB statistics both at the surface temperature of 120 K and 925 K. On the other hand, for the VASP-SRP48/FBD cases, Figures \ref{fig:BEMB}(b) and \ref{fig:BEMB925}(b) depict two important features: (i) the BE and the MB probability factors appear steeply higher magnitude in the lower frequency regime  compared to those probability factors with the EDIM; (ii) the normalized probability factor for the BE distribution is much higher in magnitude than that of the MB statistics. Although the profiles of the normalized probability factor for the BE and MB statistics are reversed by small magnitudes at higher frequency range, their contributions on the reaction probability are expected to be very low either at 120 K or 925 K. Therefore, the origin of the substantial broadening of the reaction probabilities (see Figures  \ref{fig:BEMB}(d) and \ref{fig:BEMB925}(d)) at 120 K and 925 K surface temperature with the BE statistics compared to that of the MB one in case of the VASP-SRP48/FBD could be attributed to the existence of sufficiently higher magnitude of normalized probability factor at the lower frequency range.

\begin{figure}
\centering
\includegraphics[width=\columnwidth]{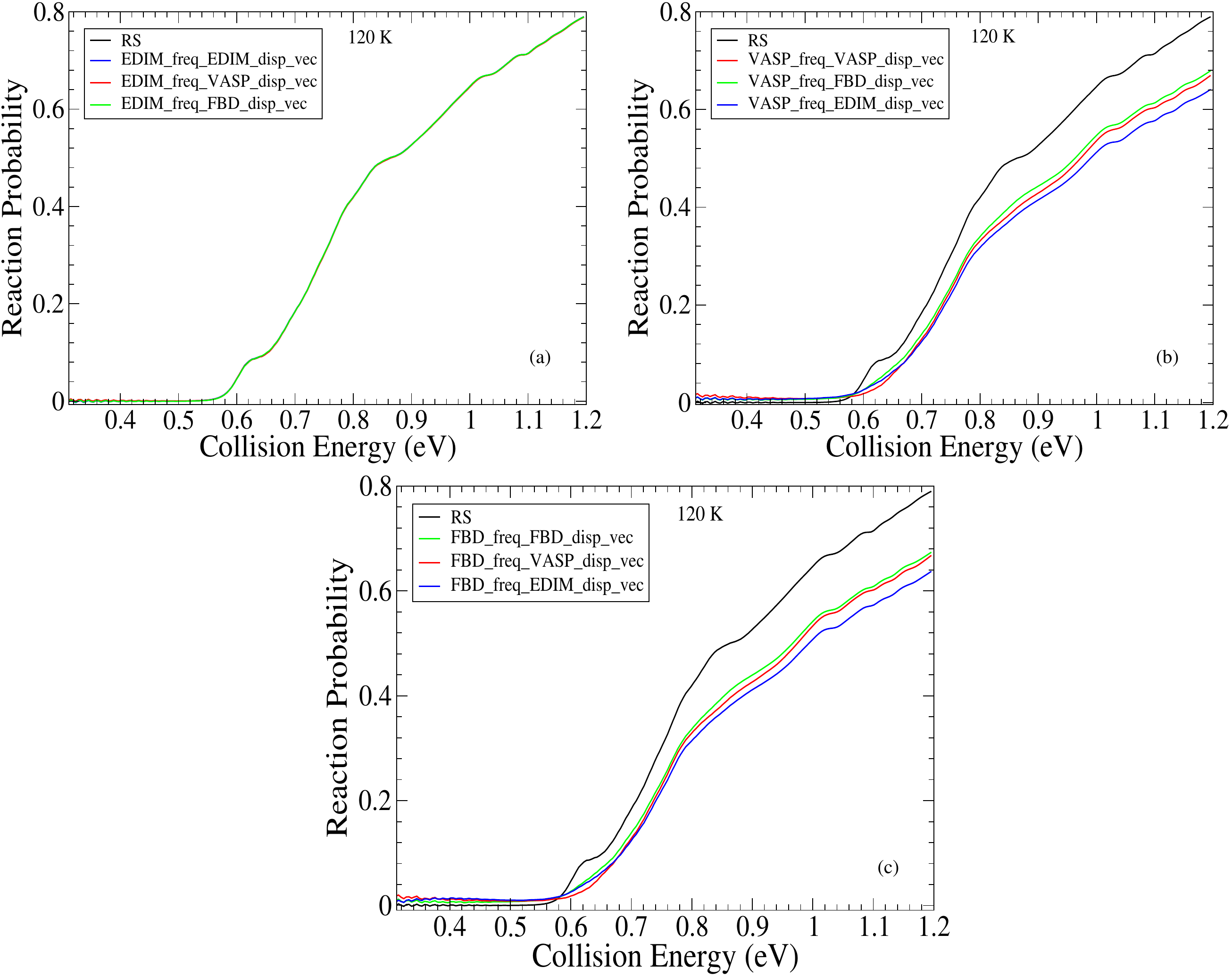}
\caption{Reaction probabilities for H$_2$ on Cu(111) calculated based on the Hartree potential at the 120 K surface temperature constructed with the (a) EDIM calculated normal mode frequencies along with the EDIM, VASP-SRP48 and FBD calculated displacement vector; (b) VASP-SRP48 calculated normal mode frequencies along with the VASP-SRP48, FBD and EDIM calculated displacement vector; (c) FBD calculated normal mode frequencies along with the FBD, VASP-SRP48 and EDIM calculated displacement vectors.}
\label{fig:cross}
\end{figure}

While exploring the effect of the normal modes on the Hartree potential, we employ a cross combination of frequencies and displacement vectors obtained from the various approaches (VASP-SRP48, FBD and EDIM) to construct a Hartree potential only with the BE probability factor and then, to calculate reaction probabilities with such potential at 120 K. In Figure \ref{fig:cross}(a), when the EDIM calculated frequencies are used along with the VASP-SRP48, FBD and EDIM calculated displacement vectors, the broadening in the reaction probability is essentially absent. On the other hand, Figure \ref{fig:cross}(b) and \ref{fig:cross}(c) show that when the VASP-SRP48 or FBD calculated normal mode frequencies are used along with the VASP-SRP48, FBD and EDIM calculated displacement vectors, the broadening in the reaction probability is substantial. Therefore, the quantization of the surface modes (see Figure \ref{fig:cross}(a)-(c)) vis-\`a-vis the distribution (see Figure \ref{fig:BEMB}(c)-(d) and \ref{fig:BEMB925}(c)-(d)) of normal modes with the BE probability factor are the key element for the origin of the broadening. 

\subsection{Influence of surface mode excitation for the reaction probability}

\begin{figure}
\centering
\includegraphics[width=\columnwidth]{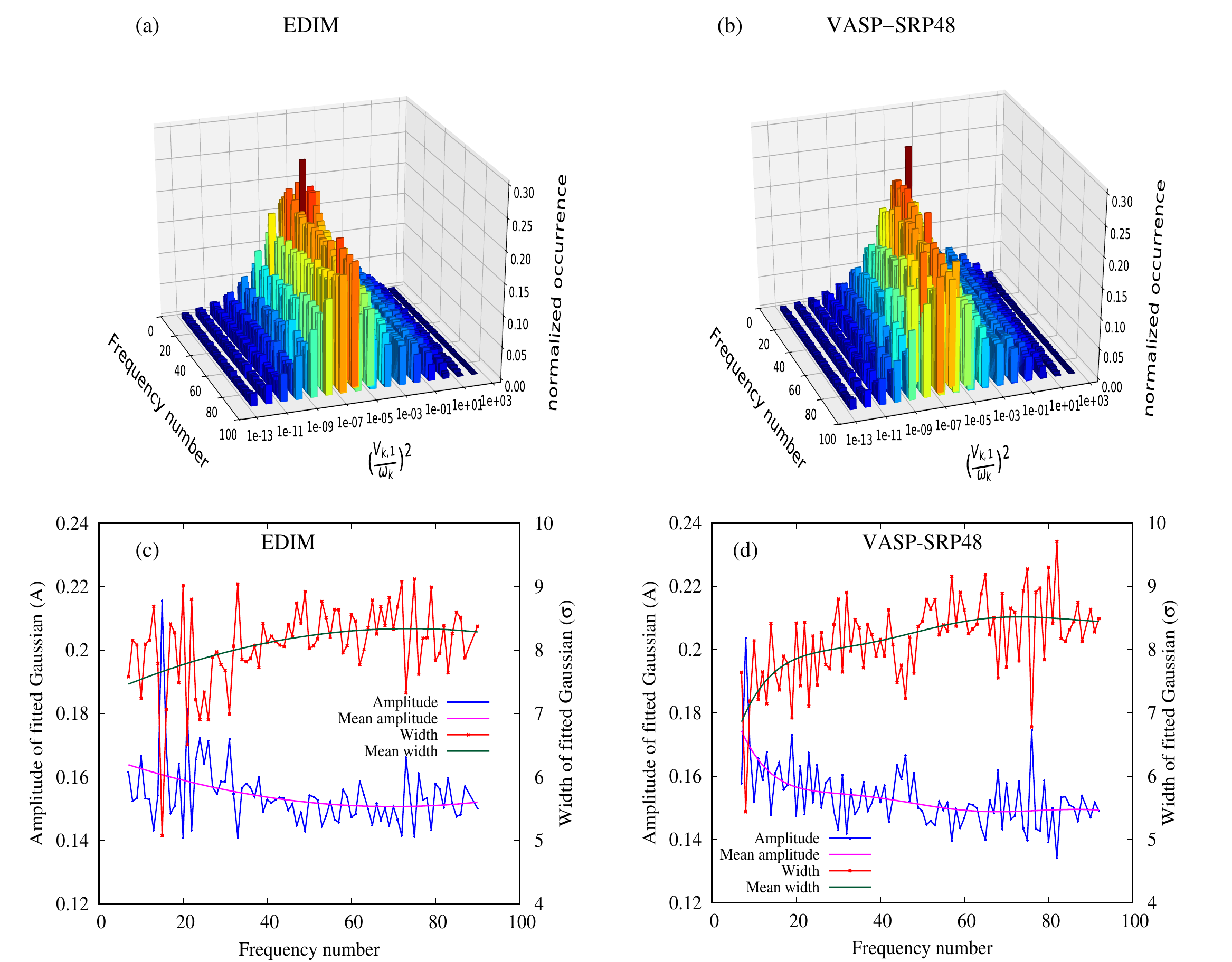}
  \caption{For the 6D SCM potential, the Normalized Occurrence of $\left(\frac{V_{k,1}}{\omega_k}\right)^2$ (=$N^\text{oc}_{V_{k,1}}$) over its various magnitudes and the frequency number ($k$) calculated from (a) the EDIM  and (b) the VASP-SRP48 surface atom interaction potential, are shown. For each vibrational mode (k), amplitudes ($A$) / mean amplitudes ($\langle A \rangle$) and widths ($\sigma$) / mean widths ($\langle \sigma \rangle$) of the fitted gaussian over the different magnitudes of $N^\text{oc}_{V_{k,1}}$ are depicted as a function of frequency number ($k$) in (c) and (d), respectively.}
\label{fig:3dhis}
\end{figure}

While constructing the Hartree potential, we need to calculate a crucially important quantity known as surface mode forcing ($\left(\frac{V_{k,1}}{\omega_k}\right)^2$), which is an average measure of surface mode excitation due to the coupling with the incoming molecule.
The contribution of $\left(\frac{V_{k,1}}{\omega_k}\right)^2$ (=$N^\text{oc}_{V_{k,1}}$) on the effective Hartree potential affecting the scattering process vis-\`a-vis reaction probability is discussed at this junction. Figure \ref{fig:3dhis}(a)-(b) display the profiles of $N^\text{oc}_{V_{k,1}}$ over each specific magnitude as function of normal mode frequency for the EDIM and VASP-SRP48 cases. The distribution of $N^\text{oc}_{V_{k,1}}$ over the different magnitudes for a specific vibrational mode (k)  has been fitted with a gaussian function ($A\exp(-(\frac{x-x_0}{\sigma})^2)$). Figure \ref{fig:3dhis}(c)-(d) depicts the variations of the amplitude ($A$)/mean amplitude ($\langle A \rangle$) and the width ($\sigma$)/mean width ($\langle \sigma \rangle$) of the fitted gaussians for the EDIM and VASP-SRP48 case, respectively as a function of the normal mode ($k$). Since the width ($\sigma$)/mean width ($\langle \sigma \rangle$) show opposite trend compared to amplitude ($A$)/mean amplitude ($\langle A \rangle$) as function of normal modes, it may not be easy to interpret the overall contribution of  $N^\text{oc}_{V_{k,1}}$ on the reaction probability. On the contrary, those profiles ($\sigma$, $\langle \sigma \rangle$, $A$, $\langle A \rangle$) for the VASP-SRP48 are steeply changing compared to the EDIM case and thereby, the VASP-SRP48 frequency spectrum affects the scattering process significantly leading to higher broadening as depicted in Figure \ref{fig:BEMB}(c)-(d), \ref{fig:BEMB925}(c)-(d) and \ref{fig:cross}(a)-(b).

\subsection{Temperature-dependent reaction and state-to-state scattering probabilities}

\begin{figure}
\centering
\includegraphics[width=\columnwidth]{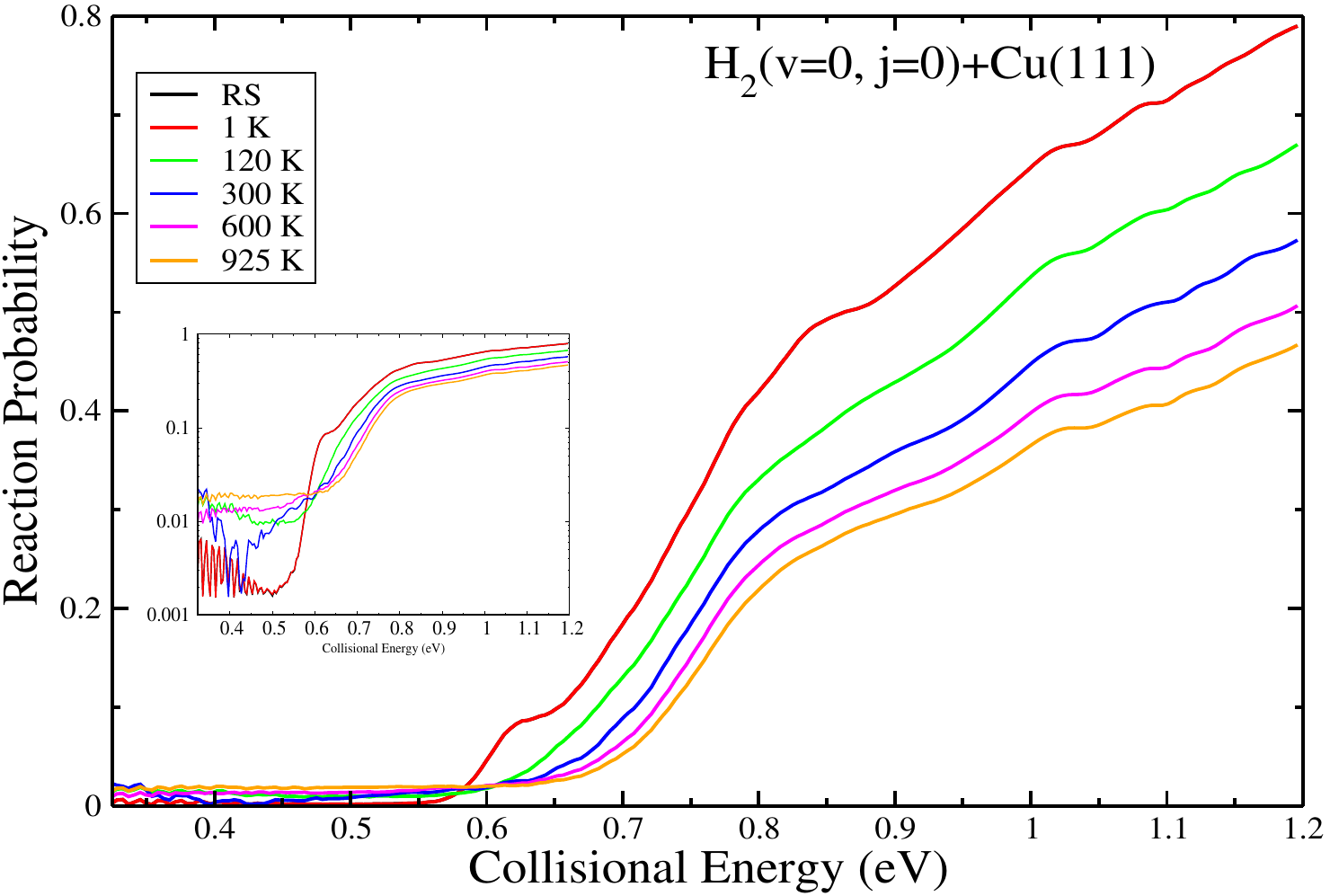}
\caption{Reaction probabilities for H$_2$ on Cu(111) based on the RS and the effective Hartree potential constructed with the VASP-SRP48 calculated normal mode frequencies at 1 K, 120 K, 300 K, 600 K and 925 K surface temperatures, where only for 600 K and 925 K, $V_{k,1}^2$ is set to zero (0) if its magnitude is $\leq$ $5.0\times 10^{-12}$ and $\leq$ $1.0\times 10^{-11}$, respectively. The reaction probabilities at different surface temperatures are also presented in log scale as inset.}
\label{fig:H2}
\end{figure}

Since the VASP-SRP48 as well as FBD calculated frequencies and displacement vectors show up substantial broadening (Figure \ref{fig:BEMB}-\ref{fig:cross}) over EDIM calculated ones, we choose VASP-SRP48 normal mode frequencies/displacement vectors to construct the Hartree potential and calculate reaction and state-to-state scattering probabilities at various surface temperatures.
For the 1 K, 120 K and 300 K surface temperatures, the dynamics are performed by considering the effective Hartree potential constructed with the VASP-SRP48 calculated normal mode frequencies without imposing any approximation, namely, the cut-off on $V_{k,1}^2$ and the converged reaction probabilities are obtained as depicted in Figure \ref{fig:H2}. On the other hand, for the 600 K and 925 K surface temperatures, it appears (numerically) that  we need to impose a cut-off on the derivative of the interaction potential ($V_{k,1}^2$) to get converged reaction probabilities (also see Figure \ref{fig:H2}), where for each mode ($k$), $V_{k,1}^2$ is set to zero if the quantity ($V_{k,1}^2$) is $\leq$ $5\times10^{-12}$ and $\leq$  $1\times10^{-11}$, respectively. There are three points to note: (a) The reaction probability profiles for the RS and the 1 K surface are perfectly merged with each other over the considered range of collisional energies (0.3 - 1.2 eV); (b) The broadening of the reaction probabilities increases with the increase of surface temperature, but the rate of broadening w.r.t. surface temperature is steadily decreasing (see Figure \ref{fig:H2}); (c) Moreover, it is evident from the log scale representation of reaction probabilities that QD results are enhanced considerably with the increase of surface temperature at low kinetic energy domain compare to the RS and the 1K ones. Such enhancement of reaction probabilities (see inset of Figure \ref{fig:H2}) may appear either due to the quantum effect at those surface temperatures or due to the numerical issues associated with the dynamical calculations, where the latter creates unphysical oscillation in QD reaction probabilities as described in the Section 2 of the supplementary material [2. \textit{(Parameters and details of the 6D QD calculations using SPO-DVR code$^{3}$)}].

Even though the inset of Figure \ref{fig:H2} reflects that at low energy region, our QD probabilities for 1 K, 120 K and 300 K surface temperatures are first diminished and then, increased after passing through minima with the increase of kinetic energy (where the positions of the minima are shifted towards low kinetic energy with the increase of surface temperature from 1 K to 120 K to 300 K), such feature is totally absent for the cases, 600 K and 925 K. On the contrary, similar feature as observed in our 1 K, 120 K, 300 K cases of theoretical calculation is found experimentally at 923$\pm$3 K by Wodtke \textit{et al.}\cite{Wodtke} due to the existence of unusual slow channel for the dissociation of H$_2$ on Cu(111) surface. These trends in 6D QD reaction probabilities at lower temperatures (1 K, 120 K and 300 K) could be arisen due to reflection related to the optical potentials or numerical inaccuracy associated with the larger time step in SPO-DVR propagation or total time propagation (or combination of all the three), while including Hartree potentials for the finite surface temperature situations.

\begin{figure}
\centering
\includegraphics[width=\columnwidth]{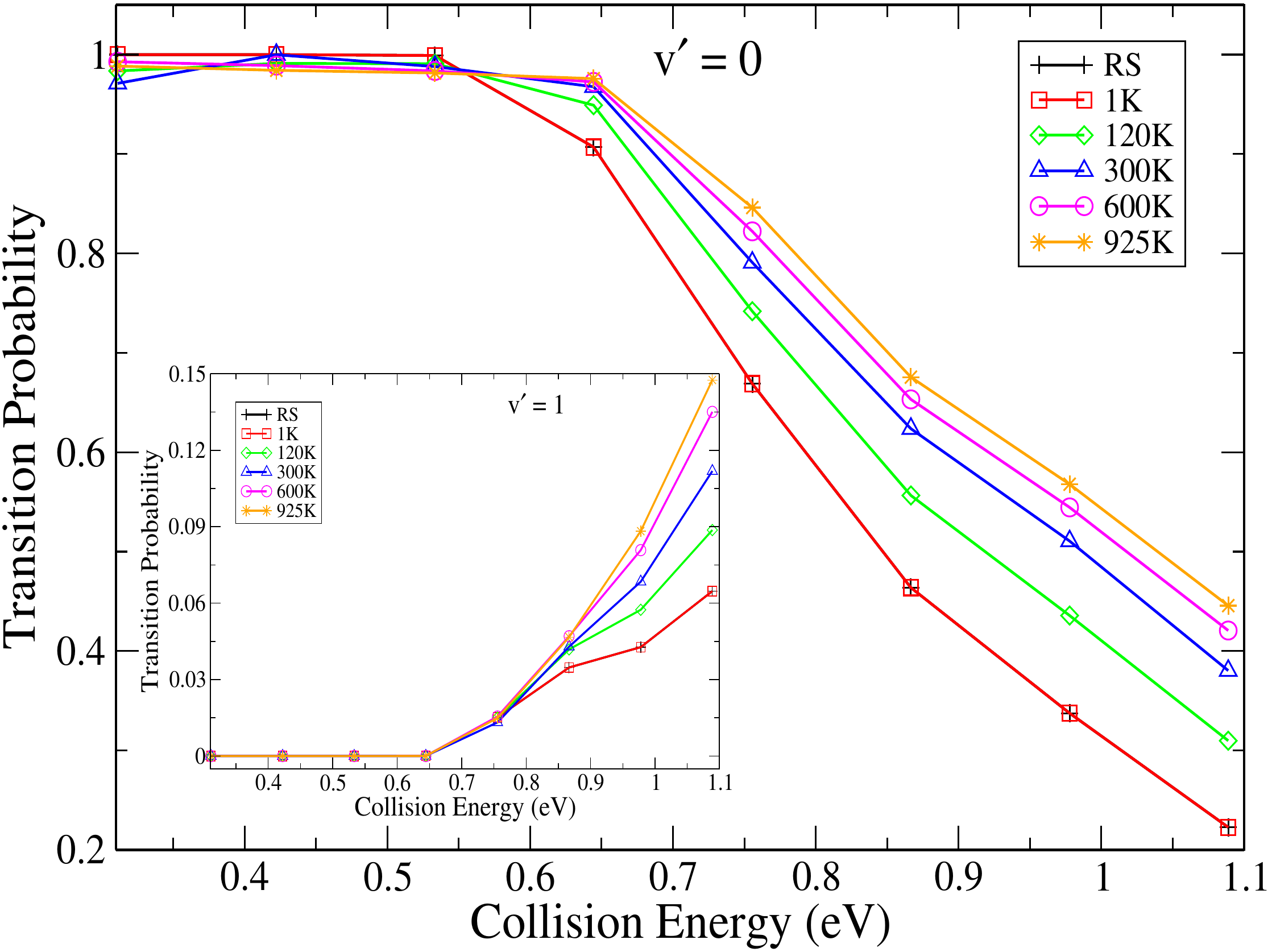}
  \caption{State-to-state transition probabilities for H$_2$($v=0,j=0$)/Cu(111) $\rightarrow$ H$_2$($v^{\prime}=0,1$)/Cu(111) on the RS and on the Hartree potential constructed with the VASP-SRP48 calculated normal mode frequencies at 1 K, 120 K, 300 K, 600 K and 925 K surface temperatures.}
\label{fig:transition}
\end{figure}

\begin{figure}
\centering
\includegraphics[width=\columnwidth]{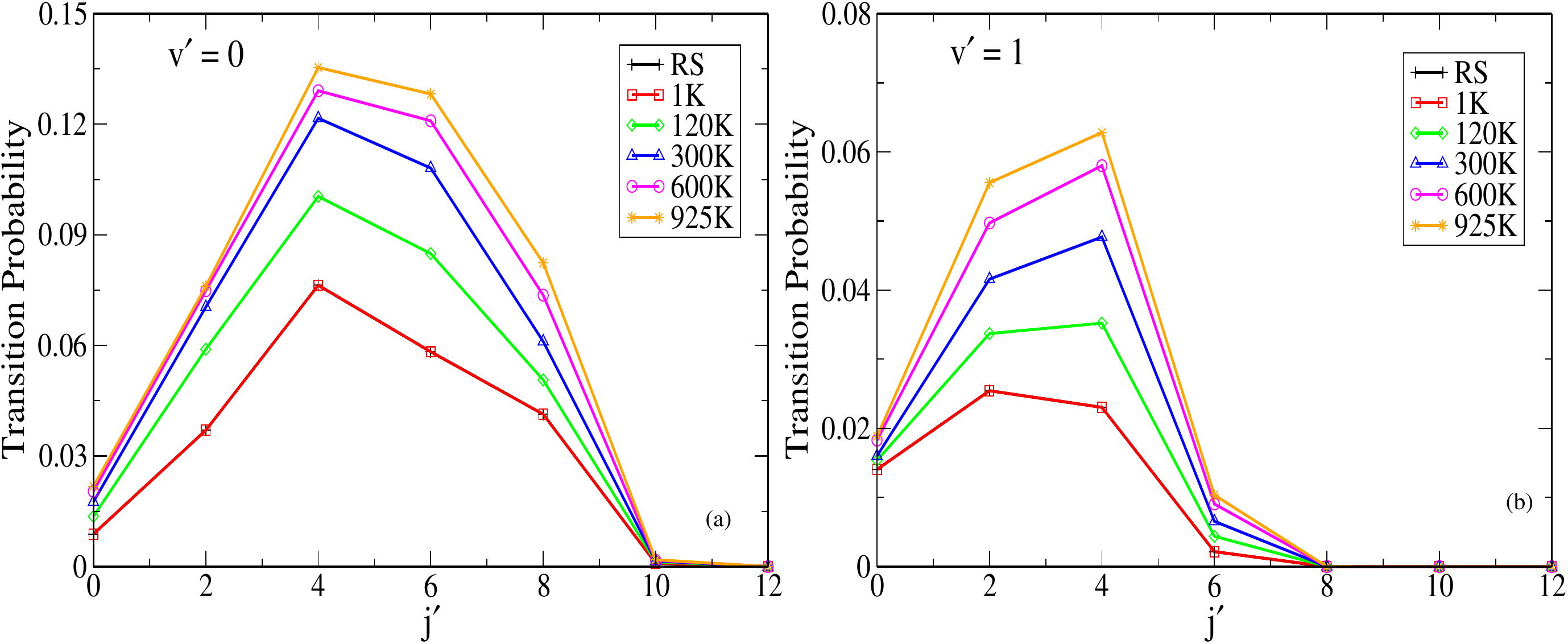}
  \caption{Final rotational state distributions for H$_2$($v=0,j=0$)/Cu(111) $\rightarrow$ H$_2$($v^{\prime}=0/1, j^{\prime}$)/Cu(111) as a function of $j^{\prime}$ at collision energy 1.08 eV on the RS and on the Hartree potential constructed with the VASP-SRP48 calculated normal mode frequencies at 1 K, 120 K, 300 K, 600 K and 925 K surface temperature.}
\label{fig:roT_state_dis}
\end{figure}

Figure \ref{fig:transition} depicts the profile of vibrational state-to-state scattering probabilities for the scattered H$_2$($v^{\prime}$ = 0, 1) molecule employing effective Hartree potential constructed with the VASP-SRP48 normal mode frequencies as a function of various initial collision energies of the incoming molecule [H$_2 (v = 0, j = 0)$] for 1 K, 120 K, 300 K, 600 K and 925 K surface temperatures. We find both the survival ($v^{\prime}$ = 0) and excitation  ($v^{\prime}$ = 1) probabilities are increasing with the increase of surface temperature and thereby, leading to a broadening of the reaction probability with the increase in surface temperature ($T_{\text{s}}$). For the VASP-SRP48 case, the final rotational state distribution is displayed in Figure \ref{fig:roT_state_dis} (a)-(b) for the scattered H$_2$($v^{\prime}=0/1, j^{\prime}$) as a function of $j^{\prime}$ at a particular collision energy (1.08 eV) for different surface temperatures. It is evident that transition probabilities attain a maximum value at a particular rotational state ($j^{\prime}$) for  all the temperatures, but the distributions become wider as the temperature increases. Moreover, the effect of temperature on the rotational state resolved transition probabilities is more pronounced in the ground vibrational state ($v^{\prime}=0$) compared to the excited state ($v^{\prime}=1$).

\section{Discussion}

\begin{figure}
\centering
\includegraphics[width=\columnwidth]{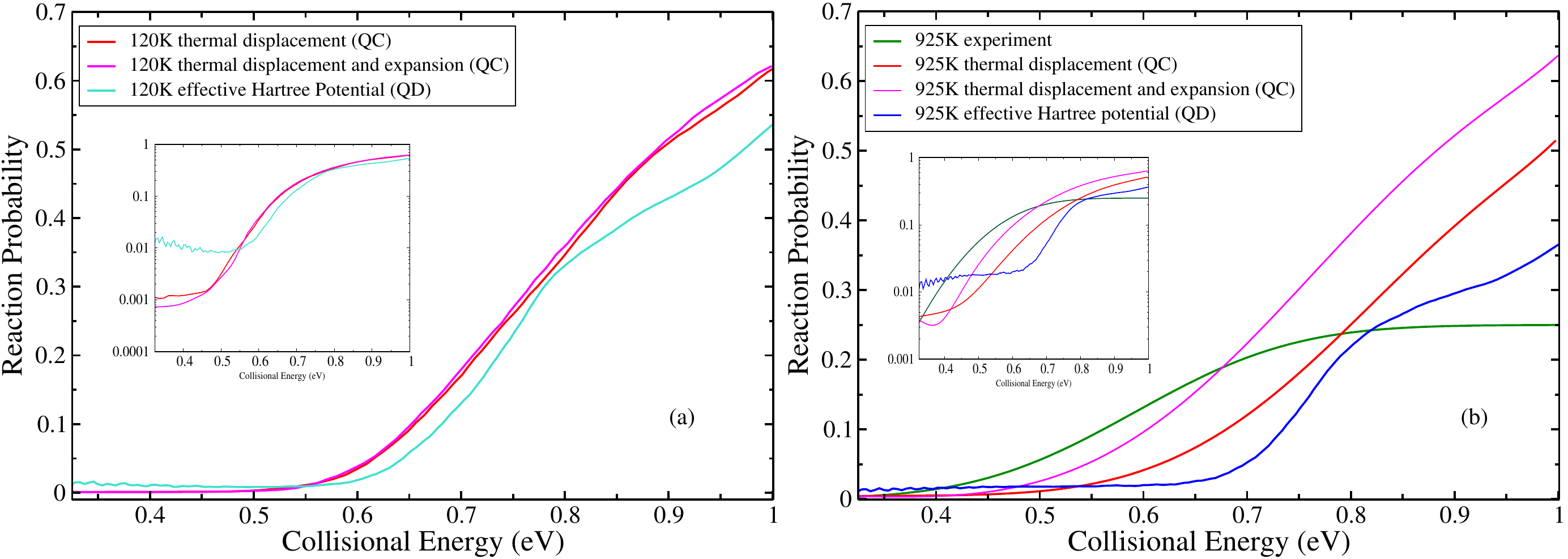}
\caption{Comparisons between the present QD profiles with (a) the various theoretical outcomes of the reaction probability at 120 K; (b) experimental\cite{CTR} (green colour) and the other theoretical results of the reaction probability at 925 K surface temperature for H$_2$ ($v=0,j=0$)-Cu(111) system. The reaction probability profile with red colour represents the QC results obtained by including the effect of thermal displacement\cite{MWMF} of surface atoms, whereas the magenta one depicts the effect of both thermal displacements as well as expansion of metal surface\cite{MWMF} on reaction probability. The turquoise and blue curve correspond to the present QD results employing the effective Hartree potential within the mean-field approximation for 120 K and 925 K surface temperatures, respectively. The reaction probabilities are also displayed in log scale as insets. The experimental data has been extracted from Ref.\cite{CTR} and is provided in Table 2 of the supplementary material.}
\label{fig:expt_theory_comp}
\end{figure}

Finally, we compare our QD results for the reaction probabilities at 120 K with other theoretical profiles and for the probabilities at 925 K both with various theoretical and experimental results. Figure \ref{fig:expt_theory_comp}(a) depicts our QD result along with QC trajectory calculations obtained from the SCM\cite{MWMF} at 120 K surface temperature. On the other hand, in Figure \ref{fig:expt_theory_comp}(b), a comparison between extracted recombinative desorption experimental\cite{CTR} data and various theoretical results has been shown for 925 K surface temperature. We emphasize that the SCM-QC\cite{MWMF} is based on the same six-dimensional VASP-SRP48 PES and includes thermal displacements of surface atoms within sudden approximation as well as expansion of the lattice at 120 K and 925 K surface temperature. Figure \ref{fig:expt_theory_comp}(a) reflects that our 6D QD-Hartree calculation provides higher reaction probability w.r.t. the other QC methods at very low collisional energy for 120 K surface temperature (also, see its inset). Although the unphysical oscillations at 120 K are much smaller compared to the RS and 1 K situations (see inset of figure 7), such enhancement of the QD reaction probabilities with respect to QC ones at low energy region could be emerged from the quantum effect or may be due to the numerical issues associated with the optical potential or larger time step or total time propagation (or combination of all the three) in SPO-DVR propagation at that temperature. On the other hand, our 6D-QD results at lower surface temperature agree quite well with other theoretically calculated reaction probabilities close to and beyond threshold energy. Whereas for higher surface temperature (925 K), it is evident from Figure \ref{fig:expt_theory_comp}(b) that at low collision energies, our theoretically estimated reaction probabilities have higher magnitude than QC ones, but over the moderate collision energies, our results deviate from the experimental and other theoretical results, which indicates the limitation of mean-field approach at higher temperature. At the same time, despite the fact that high energy domain of experimental reaction probability profile is disputable (see section 1 of supplementary material) for the H$_2$ scattering from Cu(111) in the rovibrational ground state, reaction probabilities obtained by incorporating a chemically accurate SCM potential within the mean-field approach are more close in agreement with experimental results (reported by Rettner \textit{et al.}\cite{CTR}) at those (higher) collisional energies (see Figure \ref{fig:expt_theory_comp}(b)). The substantial broadening effects at high incidence energies as observed in this 6D QD calculations could be originating due to accurate computation of surface mode frequency spectrum using VASP-SRP48 metal-metal potential, adequate description of molecule-surface interaction with SCM potential and incorporation of BE probability factor for the initial state distribution of vibrational modes involved in the configuration space.

The discrepancies between the present theoretical results and experimental observation could perhaps be reduced with the inclusion of more surface modes in the effective Hartree potential to account for the bulk properties in a realistic way. Moreover, since the present effective Hartree potential considers only the linear coupling terms, inclusion of second order molecular DOFs-surface modes correlation could improve the effect of broadening on the reaction probability profile. Regarding the accuracy of the SCM potential, even though QC calculation\cite{MWMF} using the SCM potential reproduces the AIMD sticking probabilities quite accurately, the employed normal mode configuration space with particular frequency set, displacement vectors and density of states (DOS) may also have been sampled beyond the acceptable region of the fitted SCM potential in this present QD calculation. On the other hand, the incompleteness of mean-field approach to encompass the correlations between the molecular DOFs and surface modes at higher surface temperature could be responsible for such disagreement between theory and experiment. Again, as the mean-field approach discussed here is based on a harmonic description of the surface modes, thermal lattice expansion effects are expected not to be described well.

\section{Conclusion}

In this article, we presented a formalism to take into account the role of surface vibrational modes on the reactive scattering of H$_2$ initially in its rovibrational ground state from the Cu(111) surface by considering a chemically accurate SCM potential within the mean-field approximation, where molecular DOFs are assumed to be only weakly coupled to the otherwise unaffected surface modes. A time and temperature dependent effective Hamiltonian has been constructed for linearly perturbed many oscillator model, and their initial state distribution are introduced through the BE and MB probability factors to incorporate the the effect of surface temperature. The VASP-SRP48, FBD and EDIM surface atom interaction potentials are used to calculate the characteristic surface frequency spectrum and the displacement vectors. The reaction as well as the state-resolved scattering probabilities of H$_2$ on Cu(111) initially in the rovibrational ground state are obtained by carrying out a 6D scattering calculation with the SPO-DVR code. It appears that  the distribution of initial states of normal modes with the BE probability factor  and the quantization of the surface modes are the dominating factors for the broadening in the reaction probabilities. Although we find substantial amount of broadening in reaction probability profiles with increase of surface temperature, the effect is still not close enough compared to other theoretical results and experimental observations. Such deviations could arise from five limitations: 
(a) The mean-field approach may not be theoretically accurate enough to account for all (quantum mechanical) correlations between the molecular DOFs and surface modes.
(b) Although included in the original SCM potential, the QD Hamiltonian used in this work does not account for changes in the H$_2$-Cu(111) interaction potential due to the thermal expansion of the surface lattice as at this moment it only includes the H-Cu coupling potential of the original SCM.
(c) At the same time, the sampled configuration space of normal modes incorporated may be extrapolated beyond the chemically accurate fitted domain of the SCM potential.
(d) The number of surface modes involved in the effective Hartree potential may not be sufficiently enough converged to encapsulate the actual bulk properties of the metal surface at the particular temperature. 
(e) Moreover, the surface modes-molecular DOFs coupling scheme for constructing the effective Hartree potential could be taken into account more accurately by incorporating higher order coupling terms.
(a), (b) and (c) are particularly relevant at higher surface temperatures.

\section*{Supplementary Material}
 
The supplementary material contains a brief discussion on fitting of experimental reaction probability curve, parameters for the 6D QD calculation carried out by SPO-DVR code, detail descriptions of VASP-SRP48/FBD frequency and displacement vector calculation along with their relevant parameters and comparisons among experimental and various theoretical results (presented in tabular form).

\begin{acknowledgments}
JD and SM acknowledges IACS for research fellowship. SA acknowledges DST-SERB, India, through project no. File No. CRG/2019/000793 for research funding and thanks IACS and Leiden University for access to the super-computing facility.
J.M. acknowledges financial support from the Netherlands Organisation for Scientific Research (NWO) under VIDI Grant No. 723.014.009.
\end{acknowledgments}

\clearpage

\appendix

\section{Formulation of effective Hartree potential incorporating linear coupling among molecular DOFs and surface modes}

The evolution operator for the surface modes under linear perturbation due to molecule-surface interaction is defined as $U(t, t_0)$. The  wavefunction for those surface modes at a  time $t$ can be obtained from the initial wavefunction at  time $t_0$:
\begin{eqnarray}
\Psi(t)= U(t, t_0) \Psi(t_0)
\label{Eq: Initial_wavefunction}
\end{eqnarray}

The Hartree potential that arises from the initial state $\lbrace n_0 \rbrace$ of the surface modes is defined as:
\begin{eqnarray}
\langle V \rangle_{\lbrace n_0 \rbrace} &=& \langle \Psi(t) \vert V_\text{I} \vert \Psi(t) \rangle	\nonumber  \\
&=& \langle	\Psi(t_0) \vert U^{\dagger} V_\text{I} U \vert  \Psi(t_0) \rangle \nonumber \\
&=& \langle \lbrace n_0 \rbrace \vert U^{\dagger} V_\text{I} U \vert \lbrace n_0 \rbrace \rangle \nonumber \\
&=& \sum_{\lbrace n \rbrace} \langle \lbrace n_0 \rbrace \vert U^{\dagger} V_\text{I} \vert \lbrace n \rbrace \rangle \langle \lbrace n 
\rbrace \vert 
U \vert \lbrace n_0 \rbrace \rangle \nonumber \\
&=& \sum_{\lbrace n^{\prime} \rbrace} \sum_{\lbrace n \rbrace} \langle \lbrace n_0 \rbrace \vert U^{\dagger} \vert \lbrace 
n^{\prime} \rbrace \rangle \langle \lbrace n^{\prime} \rbrace \vert V_\text{I} \vert \lbrace n \rbrace \rangle \langle \lbrace n  
\rbrace \vert U \vert \lbrace n_0  \rbrace \rangle \nonumber \\
&=& \sum_{\lbrace n^{\prime} \rbrace} \sum_{\lbrace n \rbrace} \alpha^{*}_{\lbrace n^{\prime}\rbrace \leftarrow \lbrace 
n_{0}\rbrace} (t)
\alpha_{\lbrace n \rbrace \leftarrow \lbrace n_{0}\rbrace} (t) \langle \lbrace n^{\prime} \rbrace \vert V_\text{I} \vert \lbrace n 
\rbrace \rangle.
\label{Eq: Initial_V_eff_app}
\end{eqnarray}

The amplitudes $\alpha_{\lbrace n \rbrace} (t)$ arise from the given initial state ${\lbrace n_0 \rbrace}$ as:
\begin{eqnarray}
\alpha_{\lbrace n \rbrace \leftarrow \lbrace n_{0}\rbrace} (t) = \langle \lbrace n  \rbrace \vert U \vert \lbrace n_0  \rbrace 
\rangle.
\label{Eq: amplitude_state_app}
\end{eqnarray}

As the Hartree potential is implicitly dependent upon initial state $\lbrace n_0 \rbrace$, the formulation of such a potential certainly demands an inclusion of the distribution of states rather than a specific initial state. Therefore, the effective Hartree potential is defined as:
\begin{eqnarray}
\langle V \rangle (t, T_\text{s}) = \sum_{\lbrace n_0 \rbrace} p_{\lbrace n_0 \rbrace} \langle V \rangle_{\lbrace n_0 \rbrace}
\label{Eq: state_averaged_Hartree_potential_app}
\end{eqnarray}

The distribution ($p_{\lbrace n_0 \rbrace}$) should be of the BE or the MB type:
\begin{eqnarray}
p_{\lbrace n_0 \rbrace} = \prod_{k=1}^M p_{n_k^0}^{(k)}.
\end{eqnarray}

For the quantum state ($n_k^0$) of normal mode ($\omega_k$), the BE probability factor, $p_{n_k^0}^{(k)}$ is defined as:
\begin{eqnarray}
p_{n_k^0}^{(k)} &\propto& \frac{1}{\exp \left[ \hbar \omega_k \left( n_k^0 + \frac{1}{2} \right) \beta \right] - 1 } \nonumber 
\\
&\propto& z_k^{n_k^0} \cdot z_k^{1/2} \cdot \left( 1 - z_k^{n_k^0} \cdot z_k^{1/2} \right)^{-1} \nonumber \\
&\propto& z_k^{n_k^0} \cdot z_k^{1/2} + z_k^{2 n_k^0} \cdot z_k + z_k^{3 n_k^0} \cdot z_k^{3/2} + \cdot \cdot \cdot \nonumber \\
&\propto& \sum_{q=1}^{\infty} \left( z_k^{n_k^0} \right)^q \left( z_k \right)^{\frac{q}{2}}
\label{Eq: BEP_new_factor_app}
\end{eqnarray}

and the MB probability factor, $p_{n_k^0}^{(k)}$ can be written as:
\begin{eqnarray}
p_{n_k^0}^{(k)} &\propto &   \exp \left[ - \hbar \omega_k \left( n_k^0 + \frac{1}{2} \right) \beta\right] \nonumber 
\\
&\propto& z_k^{n_k^0} \cdot z_k^{\frac{1}{2}}
\label{Eq: MBP_new_factor_app}
\end{eqnarray}
	
where $\beta = \frac{1}{k_\text{b} T_\text{s}}$ and $z_k = \exp \left(- \frac{\hbar \omega_k}{k_\text{b} T_\text{s}} \right)$. Diagonalization of the force constant (Hessian) matrix calculated from the surface atom interaction potential (VASP-SRP48, FBD and EDIM) provides the frequency set ($\lbrace \omega_k \rbrace$) of surface modes (see Figure 1).
	
The normalized probability factor for the BE or MB case is defined as: 
\begin{eqnarray}
\tilde{p}_{n_k^0}^{(k)} =  \frac{p_{n_k^0}^{(k)}}{N_{\text{BE/MB}}}, \qquad \qquad \sum_k \sum_{n_k^0 =0}^{\infty} 
\tilde{p}_{n_k^0}^{(k)} =1
\label{Eq:norm_prob}
\end{eqnarray}
where
\begin{eqnarray}
N_{\text{BE}} = \sum_k \sum_{n_k^0 =0}^{\infty} \frac{1}{\exp \left[ \hbar \omega_k \left( n_k^0 + \frac{1}{2} \right) \beta \right] 
- 1 } = \sum_k \sum_{n_k^0 =0}^{\infty}  \sum_{q=1}^{\infty} \left( z_k^{n_k^0} \right)^q \left( z_k \right)^{\frac{q}{2}}
\end{eqnarray}
and
\begin{eqnarray}
N_{\text{MB}} = \sum_k \sum_{n_k^0 =0}^{\infty}  \exp \left[ - \hbar \omega_k \left( n_k^0 + \frac{1}{2} \right) \beta\right] = \sum_k \sum_{n_k^0 =0}^{\infty} z_k^{n_k^0} \cdot z_k^{\frac{1}{2}}.
\end{eqnarray}

In order to incorporate the effect of the surface mode coupling, the interaction potential ($V_\text{I}$) among the gas molecular DOFs and surface modes can be expanded in terms of the normal mode coordinates ($Q_k$s):
\begin{eqnarray}
V_\text{I} = V_{0} + \sum_k \lambda_k V_{k,1}\, Q_k + \frac{1}{2} \sum_{kl} \gamma_{kl} V_{kl,2}\, Q_k Q_l +  \cdot \cdot \cdot,
\label{Eq: Interaction_potential_app}
\end{eqnarray}
where $V_{0}$ is the interaction potential with the lattice atoms at the equilibrium  geometry.
The first ($V_{k,1}$) and second derivatives ($V_{kl,2}$) are the cause of surface mode excitations due to diatom-surface collision and those excitations finally affect the molecular DOFs. The normal modes ($Q_k$s) are expressed in terms of boson creation ($b_k^{\dagger}$)/annihilation ($b_k$) operators such as $Q_k = A_k (b_k^{\dagger} + b_k)$ and $A_k = \sqrt{\hbar/{2\omega_k}}$. The signs of the first and second derivatives of the interaction potential have been taken into account by introducing $\lambda_k$ and $\gamma_{kl}$ as switching parameters while deriving the expression of evolution operator perturbatively.  

Considering only the linear terms, second quantized version of the interaction potential in Eq. (\ref{Eq: Interaction_potential_app}) turns into the following form:
\begin{eqnarray}
V_\text{I}=V_{0} + \sum^M_{k=1}  \lambda_k A_k(b_k F^-_k + b^+_k F^+_k) V_{k,1},
\label{Eq: Interaction_potential_Interaction_picture_app}
\end{eqnarray}
where $F^-_k = \exp(-i\omega_k t)$, $F_k^+= (F_k^-)^\ast$ are the modulatory terms associated with the boson ($b_k^{\dagger}$)/annihilation ($b_k$) operators in the interaction picture and $V_{k,1} = \partial V_\text{I}/\partial Q_k|_{eq}$.

For the BE or the MB cases, inserting Eq. (\ref{Eq: Interaction_potential_Interaction_picture_app}) in Eq. (\ref{Eq: Initial_V_eff_app}), and then in Eq. (\ref{Eq: state_averaged_Hartree_potential_app}), the effective Hartree potential becomes:
\begin{small}
\begin{eqnarray}
\langle V \rangle(t, T_\text{s}) &=& \sum_k   \lambda_k   \sum_{n^0_k} \sum_{n_k}\ \tilde{p}_{n_k^0}\ A_k\ V_{k,1} 
[n_k^{1/2}\alpha^{\ast(k)}_{n_{k-1} \leftarrow n_k^0  }(t) F_k^- + (n_k + 1)^{1/2} F_k^+ \alpha^{\ast(k)}_{n_{k+1} \leftarrow 
n_k^0 }(t)] \alpha_{n_k \leftarrow n_k^0 }^{(k)}(t)\nonumber \\
& =& \frac{1}{N_{\text{BE/MB}}}\sum_k   \lambda_k   \sum_{n^0_k} \sum_{n_k}\ p_{n_k^0}\ A_k\ V_{k,1} 
[n_k^{1/2}\alpha^{\ast(k)}_{n_{k-1} \leftarrow n_k^0  }(t) F_k^- + (n_k + 1)^{1/2} F_k^+ \alpha^{\ast(k)}_{n_{k+1} \leftarrow 
n_k^0 }(t)] \alpha_{n_k \leftarrow n_k^0 }^{(k)}(t). \nonumber \\ 
\label{Eq: Interation_picture_potential_I}
\end{eqnarray}
\end{small}

While deriving Eq. (\ref{Eq: Interation_picture_potential_I}), we used:
\begin{eqnarray}
\alpha_{\{n\}}(t)=\prod_{k=1}^M \alpha_{n_k}^{(k)}(t), \: \text{and} \: \sum_{n_k}|\alpha_{n_k}^{(k)}(t)|^2=1,
\label{Eq: alpha_prod}
\end{eqnarray}
where $\alpha_{n_k}^{(k)}(t) $ is the amplitude for the $n_k$th quantum state of the mode, $k$ obtained from the exact solution of the Linearly Forced Harmonic Oscillator (LFHO)\cite{PPJS}:
\begin{eqnarray}
\alpha_{n_k+1}^{\ast (k)}(t) = \exp(-i\beta_k - \frac{1}{2}\rho_k)[(n_k+1)! n_k^0!]^{1/2} (-i\alpha_k^-)^{n_k-n_k^0+1} 
f(\rho_k,n_k+1),
\label{Eq: alpha_n+1}
\end{eqnarray}
and
\begin{eqnarray}
\alpha_{n_k}^{(k)}(t) = \exp(i\beta_k - \frac{1}{2}\rho_k)[n_k! n_k^0!]^{1/2} (i \alpha_k^+)^{n_k-n_k^0} f(\rho_k,n_k).
\label{Eq: alpha_n}
\end{eqnarray}

The terms $\rho_k$, $\alpha_k^\pm$ and $\beta_k$ of Eqs.(\ref{Eq: alpha_n+1}) and (\ref{Eq: alpha_n}) are expressed as:
\begin{eqnarray}
\rho_k &=& \alpha_k^+ \alpha_k^-,\\
\alpha_k^\pm &=& -\frac{A_k}{\hbar} \int_{t_0}^t dt^{\prime} V_{k,1} \exp[\pm i\omega_k t^{\prime}],
\label{Eq: alpha_plus_minus}
\end{eqnarray}
and
\begin{eqnarray}
\beta_k=\frac{i}{\hbar}\int_{t_0}^t dt^{\prime} V_{k,1} \{ \exp [i\omega_k t^{\prime}]\alpha_k^
-(t^{\prime})-\exp[-i\omega_k t^{\prime}] \alpha_k^+(t^{\prime})\}.
\label{Eq: beta}
\end{eqnarray}

The $f(\rho_k,n_k)$ is written as:
\begin{eqnarray}
f(\rho_k,n_k)=\frac{1}{n_k!} L_{n_k^0}^{n_k-n_k^0} (\rho_k), \hspace{0.5 cm} n_k \geq n_k^0,
\label{Eq: laguerre_creation}
\end{eqnarray}
and
\begin{eqnarray}
f(\rho_k,n_k)=\frac{1}{n_k^0!}(-\rho_k)^{n_k^0-n_k} L_{n_k}^{n_k^0-n_k} (\rho_k), \hspace{0.5 cm} n_k < n_k^0,
\label{Eq: laguerre_annhilation}
\end{eqnarray}
where $L_{n_k}^{n_k^0-n_k}$ is the Laguerre-polynomial.

Inserting Eqs. (\ref{Eq: alpha_n+1}) - (\ref{Eq: laguerre_annhilation}) in Eq. (\ref{Eq: Interation_picture_potential_I}) and employing the BE or the MB factor for the initial distribution, we obtain
\begin{eqnarray}
\langle V \rangle (t, T_\text{s}) = \frac{1}{N_{\text{BE/MB}}}\sum_k \left( S_I^{(k)} (t, T_\text{s}) + S_{II}^{(k)} (t, T_\text{s}) \right),
\label{Eq: V_eff_SI+SII}
\end{eqnarray}
where
\begin{eqnarray}
S_I^{(k)} (t, T_\text{s}) &=& - \omega_k^{-1} \epsilon_k (t) \exp(-\rho_k) \sum_{q=1}^{\infty} \left( z_k \right)^{q/2} 
\sum_{n_k^0 = 1}^{\infty} \sum_{m=0}^{n_k^0 -1} \left( z_k^q \right)^{n_k^0} \frac{m!}{n_k^0!} \nonumber \\ &\times &
\left( \rho_k \right)^{n_k^0 - m} 
 L_m^{n_k^0 - m} (\rho_k) L_{m-1}^{n_k^0 - m +1} (\rho_k) \hspace{2cm} \text{for BE case}
\end{eqnarray}
and
\begin{eqnarray}
S_I^{(k)} (t, T_\text{s}) &=& - \omega_k^{-1} \epsilon_k (t) \exp(-\rho_k) z_k^{1/2} 
\sum_{n_k^0 = 1}^{\infty} \sum_{m=0}^{n_k^0 -1} z_k^{n_k^0} \frac{m!}{n_k^0!}\nonumber \\ &\times &
 \left( \rho_k \right)^{n_k^0 - m} L_m^{n_k^0 - m} (\rho_k) L_{m-1}^{n_k^0 - m +1} (\rho_k)  \hspace{2cm} \text{for MB case}.
\end{eqnarray}

The expression for $S_I^{(k)} (t, T_\text{s})$ can be rearranged as:
\begin{eqnarray}
S_I^{(k)} (t, T_\text{s}) &=& - \omega_k^{-1} \epsilon_k (t) \exp(-\rho_k) \sum_{q=1}^{\infty} \left( z_k \right)^{q/2}
\sum_{n=0}^{\infty} \sum_{m=0}^{\infty} \left( z_k^q \right)^{m+n} \frac{n!}{\left( n+m \right)!} \nonumber \\ &\times &
\left( \rho_k \right)^m L_n^m 
(\rho_k) L_{n-1}^{m+1} (\rho_k) \hspace{4cm} \text{for BE case} 
\label{Eq: S_I}
\end{eqnarray}
and
\begin{eqnarray}
S_I^{(k)} (t, T_\text{s}) &=& - \omega_k^{-1} \epsilon_k (t) \exp(-\rho_k) z_k^{1/2}
\sum_{n=0}^{\infty} \sum_{m=0}^{\infty} z_k^{m+n} \frac{n!}{\left( n+m \right)!} \nonumber \\ &\times &
\left( \rho_k \right)^m L_n^m (\rho_k) L_{n-1}^{m+1} (\rho_k) \hspace{4cm} \text{for MB case}.
\label{Eq: S_I_MBP}
\end{eqnarray}

Similarly, the second term $S_{II}^{(k)} (t, T_\text{s})$ in Eq. (\ref{Eq: V_eff_SI+SII}) can be expressed as:
\begin{eqnarray}
S_{II}^{(k)} (t, T_\text{s}) &=& \omega_k^{-1} \epsilon_k (t) \exp(-\rho_k) \sum_{q=1}^{\infty} \left( z_k \right)^{q/2} 
\sum_{n_k^0 = 0}^{\infty} \sum_{m=n_k^0}^{\infty} \left( z_k^q \right)^{n_k^0} \frac{n_k^0!}{m!}\nonumber \\ &\times & \left( \rho_k \right)^{m-n_k^0} 
L_m^{m-n_k^0} (\rho_k) L_m^{m-n_k^0+1} (\rho_k)\hspace{2cm} \text{for BE case} 
\end{eqnarray}
and
\begin{eqnarray}
S_{II}^{(k)} (t, T_\text{s}) &=& \omega_k^{-1} \epsilon_k (t) \exp(-\rho_k) z_k^{1/2} 
\sum_{n_k^0 = 0}^{\infty} \sum_{m=n_k^0}^{\infty} z_k^{n_k^0} \frac{n_k^0!}{m!} \nonumber \\ &\times & \left( \rho_k \right)^{m-n_k^0} L_m^{m-n_k^0} (\rho_k) L_m^{m-n_k^0+1} (\rho_k), \hspace{2cm} \text{for MB case} 
\end{eqnarray}
which may be rewritten as:
\begin{eqnarray}
S_{II}^{(k)} (t, T_\text{s}) &=& \omega_k^{-1} \epsilon_k (t) \exp(-\rho_k)  \sum_{q=1}^{\infty} \left( z_k \right)^{q/2} 
\sum_{n=0}^{\infty} \sum_{m=0}^{\infty} \left( z_k^q \right)^n \frac{n!}{(n+m)!} \nonumber \\ &\times & \left( \rho_k \right)^m L_n^{m} (\rho_k) 
L_n^{m+1} (\rho_k)  \hspace{4cm} \text{for BE case}
\label{Eq: S_II}
\end{eqnarray}
and
\begin{eqnarray}
S_{II}^{(k)} (t, T_\text{s}) &=& \omega_k^{-1} \epsilon_k (t) \exp(-\rho_k)  \sum_{q=1}^{\infty} z_k^{1/2} 
\sum_{n=0}^{\infty} \sum_{m=0}^{\infty} z_k^n \frac{n!}{(n+m)!} \nonumber \\ &\times & 
\left( \rho_k \right)^m L_n^{m} (\rho_k) 
L_n^{m+1} (\rho_k)  \hspace{4cm} \text{for MB case}. 
\label{Eq: S_II_MBP}
\end{eqnarray}

In Eq. (\ref{Eq: S_I}) - (\ref{Eq: S_II_MBP}), the explicit time dependent quantity, $\epsilon_k (t)$ is defined as:
\begin{eqnarray}
\epsilon_k(t) = \lambda_k V_{k,1} \int_{t_0}^{t} dt^{\prime} V_{k,1} \sin \left[ \omega_k \left( t^{\prime} 
- t \right) \right].
\label{Eq: epsilon}
\end{eqnarray} 

Since the first derivative of the interaction potential ($V_{k,1}$) is time independent, it can be taken out of the integral and the integration over time can be performed analytically. 

Applying Eq. (\ref{Eq: S_I}) and Eq. (\ref{Eq: S_II}) on Eq.(\ref{Eq: V_eff_SI+SII}), the form of effective potential has been turned into the following form for the BE case:
\begin{eqnarray}
 \langle V \rangle (t, T_\text{s})& =&\frac{1}{N_\text{BE}}\sum_k \omega_k^{-1} \epsilon_k (t) \exp(- \rho_k) \nonumber \\
& \times & \bigg\{ \sum_{q=1}^{\infty} \left( z_k 
\right)^{q/2} \sum_{n=0}^{\infty} \sum_{m=0}^{\infty} \left( z_k^q  \right)^{n+m} \frac{n!}{(n+m)!}
\left( \rho_k \right)^m L_n^{m} (\rho_k) L_n^{m} (\rho_k) \nonumber \\
& + &\sum_{q=1}^{\infty} \left( z_k \right)^{q/2} \sum_{n=0}^{\infty} \sum_{m=0}^{\infty} \left( z_k^q  \right)^{n} \frac{n!}
{(n+m)!}
\left( \rho_k \right)^m L_n^{m} (\rho_k) L_n^{m} (\rho_k) \nonumber \\
& +& \sum_{q=1}^{\infty} \left( z_k \right)^{q/2} \sum_{n=0}^{\infty} \sum_{m=0}^{\infty} \left( z_k^q  \right)^{n} \frac{n!}
{(n+m)!}
\left( \rho_k \right)^m L_n^{m} (\rho_k) L_{n-1}^{m} (\rho_k) \nonumber \\
&-& \sum_{q=1}^{\infty} \left( z_k \right)^{q/2} \sum_{n=0}^{\infty} \sum_{m=0}^{\infty} \left( z_k^q  \right)^{n+m} \frac{n!}
{(n+m)!}
\left( \rho_k \right)^m L_n^{m} (\rho_k) L_n^{m+1} (\rho_k) \bigg\}.
\label{Eq: V_eff_pre_reduced}
\end{eqnarray}

Similarly, the expression of the effective potential for the MB factor is:
\begin{eqnarray}
 \langle V \rangle (t, T_\text{s}) &=&\frac{1}{N_{\text{MB}}}\sum_k \omega_k^{-1} \epsilon_k (t) \exp(- \rho_k) \nonumber \\
&  \times & \bigg\{ z_k^{1/2} \sum_{n=0}^{\infty} \sum_{m=0}^{\infty} z_k^{n+m} \frac{n!}{(n+m)!}
\left( \rho_k \right)^m L_n^{m} (\rho_k) L_n^{m} (\rho_k) \nonumber \\
& + & z_k^{1/2} \sum_{n=0}^{\infty} \sum_{m=0}^{\infty} z_k^{n} \frac{n!}
{(n+m)!}
\left( \rho_k \right)^m L_n^{m} (\rho_k) L_n^{m} (\rho_k) \nonumber \\
& + &z_k^{1/2} \sum_{n=0}^{\infty} \sum_{m=0}^{\infty} z_k^{n} \frac{n!}
{(n+m)!}
\left( \rho_k \right)^m L_n^{m} (\rho_k) L_{n-1}^{m} (\rho_k) \nonumber \\
&-& z_k^{1/2} \sum_{n=0}^{\infty} \sum_{m=0}^{\infty} z_k^{n+m} \frac{n!}
{(n+m)!}
\left( \rho_k \right)^m L_n^{m} (\rho_k) L_n^{m+1} (\rho_k) \bigg\}.
\label{Eq: V_eff_pre_reduced_MBP}
\end{eqnarray}

The following identity:
\begin{eqnarray}
\sum_{p=0}^\infty p! \ L_p^\alpha(x) L_p^\alpha(y) \frac{z^p}{(p+\alpha)!}
=\bigg[\frac{(xyz)^{-\frac{\alpha}{2}}}{(1-z)}\bigg] \exp \bigg[-\frac{z(x+y)}{(1-z)}\bigg]\times I_\alpha [2(xyz)^
\frac{1}{2}/(1-z)],\nonumber
\end{eqnarray} 
has been plugged in Eq. (\ref{Eq: V_eff_pre_reduced}) and (\ref{Eq: V_eff_pre_reduced_MBP}) to arrive at a more simplified form for both the BE and the MB cases, respectively:
\begin{eqnarray}
\langle V \rangle(t, T_\text{s})&=& \frac{1}{N_{\text{BE}}} \sum_k \omega_k^{-1} \epsilon_k (t) \sum_{q=1}^\infty \frac{z_k^{q/2}}{(1-
z_k^q)} 
\exp{[\rho_k (1+z_k^q)/(z_k^q-1)]} \nonumber\\
&\times& \{S^{q+}_k(t, T_\text{s}) + S^{q-}_k(t, T_\text{s}) -I_0(t^q_k)\}, \hspace{2cm} \text{for BE case}
\label{Eq: V_eff_reduced}
\end{eqnarray}
and
\begin{eqnarray}
\langle V \rangle(t, T_\text{s})&=& \frac{1}{N_{\text{MB}}} \sum_k \omega_k^{-1} \epsilon_k (t) \frac{z_k^{1/2}}{(1-z_k)} 
\exp{[\rho_k (1+z_k)/(z_k-1)]} \nonumber\\
&\times& \{S^{+}_k(t, T_\text{s}) + S^{-}_k(t, T_\text{s}) -I_0(t_k)\}, \hspace{2cm} \text{for MB case}
\label{Eq: V_eff_reduced_MBP}
\end{eqnarray}
where the terms $S_k^{q \pm}(t, T_\text{s})$ can be expressed as $S_k^{q \pm}(t, T_\text{s}) = \displaystyle{\sum_{m=0}^\infty} (z_k^q)^{\pm m/2} I_m(t_k^q)$ by using the modified Bessel-function of the first kind $I_m(t_k^q)$, and $t_k^q = 2 \rho_k (z_k^q)^{1/2}/(1-z_k^q)$.

As $S_k^{q+}(t, T_\text{s}) + S_k^{q-}(t, T_\text{s}) - I_0(t_k^q) = \exp[\rho_k(1+z_k^q)/(1-z_k^q)]$, ($q$ = 1, 2, 3, ....), the
expression in Eq. (\ref{Eq: V_eff_reduced}) and (\ref{Eq: V_eff_reduced_MBP}) are further simplified to:
\begin{eqnarray}
\langle V \rangle(t, T_\text{s}) &=& \frac{1}{N_{\text{BE}}} \sum_k \omega_k^{-1} \epsilon_k (t)  \sum^{\infty}_{q=1}\frac{z_k^{q/2}}{(1-
z^q_k)}, \hspace{1cm} \text{for BE case}
 \label{Eq: V_eff_final}
\end{eqnarray}
and
\begin{eqnarray}
\langle V \rangle(t, T_\text{s}) &=& \frac{1}{N_{\text{MB}}} \sum_k \omega_k^{-1} \epsilon_k (t)  \frac{z_k^{1/2}}{(1-
z_k)}. \hspace{1cm} \text{for MB case}
 \label{Eq: V_eff_final_MBP}
\end{eqnarray}
Thus, the final form of the effective Hartree potential for the BE case is written as:
\begin{eqnarray}
V_\text{eff}^\text{BE} (R, \theta, \phi, X, Y, Z, t, T_\text{s}) = \frac{1}{N_{\text{BE}}} \sum_{k=7}^{3N} \lambda_k \frac{1}{\omega_k^2} V_{k,1}^2 
\left[ \cos \omega_k (t - t_0) - 1 \right]\
 \sum^{\infty}_{q=1}\frac{z_k^{q/2}}{(1-z^q_k)} 
\label{Eq: V_Hartree_final_form}
\end{eqnarray}
and for the MB case:
\begin{eqnarray}
V_\text{eff}^\text{MB} (R, \theta, \phi, X, Y, Z, t, T_\text{s}) = \frac{1}{N_{\text{MB}}} \sum_{k=7}^{3N} \lambda_k \frac{1}{\omega_k^2} V_{k,1}^2 
\left[ \cos \omega_k (t - t_0) - 1 \right]\
\frac{z_k^{1/2}}{(1-z_k)} 
\label{Eq: V_Hartree_final_form_MBP}
\end{eqnarray}
The normalization for the BE or the MB cases can be simplified as:
\begin{eqnarray}
N_{\text{BE}} = \sum_k \sum_{n_k^0 =0}^{\infty}  \sum_{q=1}^{\infty} \left( z_k^{n_k^0} \right)^q \left( z_k \right)^{\frac{q}{2}} &=& \sum_k \sum_{q=1}^{\infty} \left( z_k \right)^{\frac{q}{2}} \sum_{n_k^0 =0}^{\infty} \exp \left[ - \frac{q n_k^0 \hbar 
 \omega_k}{k_\text{b} T_\text{s}} \right] \nonumber \\
&=& \sum_k \sum_{q=1}^{\infty} \frac{\left( z_k \right)^{\frac{q}{2}}}{\left( 1 - z_k^q \right)}
\label{Eq: final_form_of_norm}
\end{eqnarray}
and
\begin{eqnarray}
N_{\text{MB}} = \sum_k \sum_{n_k^0 =0}^{\infty} z_k^{n_k^0} \cdot z_k^{\frac{1}{2}}
 &=& \sum_k z_k^{\frac{1}{2}} \sum_{n_k^0 =0}^{\infty} \exp \left[ - \frac{n_k^0 \hbar 
 \omega_k}{k_\text{b} T_\text{s}} \right] \nonumber \\
&=& \sum_k \frac{z_k^{\frac{1}{2}}}{\left( 1 - z_k\right)}
\label{Eq: final_form_of_norm_MBP}
\end{eqnarray}

\section{Evolution of first derivative of the interaction potential w.r.t. metal atom position $\left(\dfrac{\partial V_{a\alpha}^\text{Cu-H}}{\partial X_{\alpha i}}\right)$}
The interaction potential between a metal (Cu) atom of the surface and a gas atom of the molecule
has been written as\cite{MWMF}:
\begin{eqnarray}
V^\text{Cu-H}_{a\alpha}(r_{a\alpha})=(1-\rho(r_{a\alpha}))V_\text{Ryd}(r_{a\alpha})+\rho(r_{a\alpha})V_\text{Ryd}(b_{2})
\label{totpot}
\end{eqnarray}
where $V_\text{Ryd}(r_{a\alpha})$ can be defined as,
\begin{equation}
V_\text{Ryd}(r_{a\alpha})=-\exp\lbrace-l(r_{a\alpha}-z)\rbrace\sum^{3}_{k=0}(c_{k}(r_{a\alpha}-z)^{k})
\label{RydPot}
\end{equation} 
and the $\rho(r_{a\alpha})$ is expressed as,
\begin{eqnarray}
\rho(r_{a\alpha}) =
  \begin{cases}
    0       & \quad \text{if } r_{a\alpha} < b_{2}\\
    \dfrac{1}{2}\cos\left(\dfrac{\pi(r_{a\alpha}-b_{2})}{b_{2}-b_{1}}\right) + \dfrac{1}{2}  & \quad \text{if }  b_{1}\leq 
    r_{a\alpha} \leq b_{2}\\
    1       & \quad \text{if } r_{a\alpha} > b_{2}
  \end{cases}
\label{rho}
\end{eqnarray}

The interaction potential ($V^\text{Cu-H}_{a\alpha}(r_{a\alpha})$) as displayed in Eq. (\ref{totpot})-(\ref{rho}) is a function of the gas-metal distance ($r_{a\alpha}$) with Rydberg parameters ($b_{1}$, $b_{2}$, $c_{0}$, $c_{1}$, $c_{2}$, $c_{3}$, $l$, $z$ ). Those parameters ($P_i$) are dependent on the two quantities
$P_{i,\text{I}}$ and $P_{i,\text{II}}$, which  are related to the H-H separation ($R$) and a pure two-body, H-Cu ($r_{a\alpha}$) component of three-body SCM potential, respectively. 
\begin{eqnarray}
P_i =
  \begin{cases}
    P_{i,\text{I}} R_\text{min} +   P_{i,\text{II}}      & \quad \text{if } R < R_\text{min}\\
   P_{i,\text{I}} R +   P_{i,\text{II}}   & \quad \text{if }  R_\text{min} \leq R \leq R_\text{max}     \\
    P_{i,\text{I}} R_\text{max} +   P_{i,\text{II}}     & \quad \text{if } R > R_\text{max}
  \end{cases}
\label{parameters}
\end{eqnarray}
The first derivative of the interaction potential is given by:
\begin{eqnarray}
\left[\dfrac{\partial V_{a\alpha}^\text{Cu-H}(r_{a\alpha}^\text{id})}{\partial X_{\alpha i}}\right]=0
\end{eqnarray}
and
\begin{small}
\begin{eqnarray}
\dfrac{\partial V_{a\alpha}^\text{Cu-H}(r_{a\alpha})}{\partial X_{\alpha i}} =
  \begin{cases}
   \Big[-lV_\text{Ryd}(r_{a\alpha})-\exp
\lbrace-l(r_{a\alpha}-z)\rbrace \times \\  \Big(c_{1} + 2c_{2}(r_{a\alpha}-z)+3c_{3}(r_{a\alpha}-z)^{2}\Big)\Big]
\left(\frac{X_{\alpha i}-X_{a i}}
{r_{a\alpha}}\right)      & \quad  \text{if } r_{a\alpha} < b_{2}\\ \\
 \Big[\frac{1}{2} \sin
 \left(\frac{\pi(r_{a\alpha}-b_2)}{b_2-b_1}\right)  \left(\frac{\pi}{b_2-b_1} \right) \Big(V_\text{Ryd}(r_{a\alpha}) -V_\text{Ryd}(b_2)
  \Big)+ \\ (1-\rho(r_{a\alpha}))   \Big\{-lV_\text{Ryd}(r_{a\alpha})-\exp
\lbrace-l(r_{a\alpha}-z)\rbrace \times \\ \Big(c_{1} + 2c_{2}(r_{a\alpha}-z)+3c_{3}(r_{a\alpha}-z)^{2}\Big)\Big\}\Big]
 \Big(\frac{X_{\alpha i}-X_{a i}}
{r_{a\alpha}}\Big)  & \quad  \text{if }  b_{1}\leq 
    r_{a\alpha} \leq b_{2}\\ \\
    0       & \quad  \text{if } r_{a\alpha} > b_{2}
  \end{cases}
\label{dVdX}
\end{eqnarray}
\end{small}

\subsection{Evolution of \quad $\left[\dfrac{\partial V_{a\alpha}^\text{Cu-H}(r_{a\alpha}^\text{id})}{\partial X_{\alpha i}}\right]$}

\begin{eqnarray}
r_{a\alpha}^\text{id} &=& \sqrt{\sum_{i}(X_{a i} - X_{\alpha i}^\text{id})^{2}} \nonumber \\
\dfrac{\partial r_{a\alpha}^\text{id}}{\partial X_{\alpha i}}&=&0 \nonumber \\ \\
\therefore \left[\dfrac{\partial V_{a\alpha}^\text{Cu-H}(r_{a\alpha}^\text{id})}{\partial X_{\alpha i}}\right]&=&\dfrac{\partial V_{a\alpha}^\text{Cu-H}
(r_{a\alpha}^\text{id})}{\partial r_{a\alpha}^\text{id}}.\dfrac{\partial r_{a\alpha}^\text{id}}{\partial X_{\alpha i}}=0\nonumber\\
\end{eqnarray}

\subsection{Evolution of \quad $\left[\dfrac{\partial V_{a\alpha}^\text{Cu-H}(r_{a\alpha})}{\partial X_{\alpha i}}= \dfrac{\partial V_{a\alpha}^\text{Cu-H}(r_{a\alpha})}{\partial r_{a \alpha}} \cdot \dfrac{\partial r_{a\alpha}}{\partial X_{\alpha i}}\right]$}

\begin{eqnarray}
\frac{\partial r_{a\alpha}}{\partial X_{\alpha i}}&=&\dfrac{X_{\alpha i} - X_{a i}}{r_{a\alpha}} \\ \nonumber \\
\frac{\partial V^\text{Cu-H}_{a\alpha}(r_{a\alpha})}{\partial r_{a \alpha}} &=& \frac{\partial}{\partial r_{a\alpha}} \Big[(1-
\rho(r_{a\alpha}))V_\text{Ryd}(r_{a\alpha}) +\rho(r_{a\alpha}) V_\text{Ryd}(b_2) \Big]\nonumber \\
&=& - \frac{\partial \rho(r_{a\alpha})}{\partial r_{a\alpha}}V_\text{Ryd} (r_{a\alpha}) + (1-\rho(r_{a \alpha})) \frac{\partial 
V_\text{Ryd}(r_{a\alpha})}{\partial r_{a \alpha}} \nonumber \\
&& +\frac{\partial \rho(r_{a\alpha})}{\partial r_{a\alpha}}V_\text{Ryd} (b_2)
 +\rho(r_{a\alpha})\frac{\partial V_\text{Ryd} (b_2)}{\partial r_{a\alpha}}
\end{eqnarray}
where $r_{a\alpha} =\sqrt{\sum_{i}(X_{a i} - X_{\alpha i})^{2}}$ and $\rho(r_{a\alpha})\frac{\partial V_\text{Ryd} (b_2)}{\partial r_{a\alpha}}$ = 0 as $V_\text{Ryd} (b_2)$ is independent of $r_{a\alpha}$.

\subsubsection{If $r_{a\alpha}$ $<$ $b_{2}$:}

In this range, $\rho(r_{a\alpha})$ = 0 \; and \; $\frac{\partial \rho (r_{a\alpha})}{\partial r_{a\alpha}}$ = 0.
\begin{eqnarray}
\frac{\partial V^\text{Cu-H}_{a\alpha}(r_{a\alpha})}{\partial r_{a \alpha}} &=& \frac{\partial 
V_\text{Ryd}(r_{a\alpha})}{\partial r_{a \alpha}}\nonumber \\
&=&\dfrac{\partial}{\partial r_{a \alpha}}\Big[-\exp\lbrace-l(r_{a\alpha}-z)\rbrace\sum^{3}_{k=0}(c_{k}(r_{a\alpha}-
z)^{k})\Big]\nonumber\\
&=&-lV_\text{Ryd}(r_{a\alpha})-\exp
\lbrace-l(r_{a\alpha}-z)\rbrace\left[c_{1} + 2c_{2}(r_{a\alpha}-z)+3c_{3}(r_{a\alpha}-z)^{2}\right]\nonumber\\
\end{eqnarray}

\begin{eqnarray}
\therefore \dfrac{\partial V_{a\alpha}^\text{Cu-H}(r_{a\alpha})}{\partial r_{a\alpha}}.\dfrac{\partial r_{a\alpha}}{\partial X_{\alpha 
i}}&=&\Big[-lV_\text{Ryd}(r_{a\alpha})-\exp
\lbrace-l(r_{a\alpha}-z)\rbrace\left[c_{1} + 2c_{2}(r_{a\alpha}-z)+3c_{3}(r_{a\alpha}-z)^{2}\right]\Big]\nonumber\\
&&\times \left(\frac{X_{\alpha i}-X_{a i}}
{r_{a\alpha}}\right)
\label{1st1st}
\end{eqnarray}

\subsubsection{If $b_{1}$ $\leq$ $r_{a\alpha}$ $\leq$  $b_{2}$:}

\text{In this range,}
 $\rho(r_{a\alpha}) = \frac{1}{2} \cos \left(\frac{\pi(r_{a\alpha}-b_2)}{b_2-b_1}\right) + 
\frac{1}{2}$ \; and \;
$\frac{\partial \rho (r_{a\alpha})}{\partial r_{a\alpha}} = -\frac{1}{2}
 \sin \left(\frac{\pi(r_{a\alpha}-b_2)}{b_2-b_1}\right)\frac{\pi}{b_2-b_1}$.
\begin{eqnarray}
\frac{\partial V^\text{Cu-H}_{a\alpha}(r_{a\alpha})}{\partial r_{a \alpha}} &=&\frac{1}{2}
 \sin \left(\frac{\pi(r_{a\alpha}-b_2)}{b_2-b_1}\right)\left(\dfrac{\pi}{b_2-b_1}\right)V_\text{Ryd} (r_{a\alpha}) 
+(1-\rho(r_{a\alpha})) \frac{\partial V_\text{Ryd}(r_{a\alpha})}{\partial r_{a\alpha}} \nonumber\\
&&-\frac{1}{2}
 \sin \left(\frac{\pi(r_{a\alpha}-b_2)}{b_2-b_1}\right)\left(\dfrac{\pi}{b_2-b_1}\right)V_\text{Ryd} (b_{2})\nonumber\\
&=& \frac{1}{2}
 \sin \left(\frac{\pi(r_{a\alpha}-b_2)}{b_2-b_1}\right)\left(\dfrac{\pi}{b_2-b_1}\right) \left( V_\text{Ryd} (r_{a\alpha}) -V_\text{Ryd}
 (b_2) \right) 
 + (1-\rho(r_{a\alpha}))\nonumber\\
 &&\Big[-lV_\text{Ryd}(r_{a\alpha})-\exp
\lbrace-l(r_{a\alpha}-z)\rbrace\left[c_{1} + 2c_{2}(r_{a\alpha}-z)+3c_{3}(r_{a\alpha}-z)^{2}\right]\Big],\nonumber \\
\end{eqnarray}
where $\frac{\partial V_\text{Ryd}(r_{a\alpha})}{\partial r_{a \alpha}} = -lV_\text{Ryd}(r_{a\alpha})-\exp
\lbrace-l(r_{a\alpha}-z)\rbrace\left[c_{1} + 2c_{2}(r_{a\alpha}-z)+3c_{3}(r_{a\alpha}-z)^{2}\right]$.

\begin{eqnarray}
\therefore \dfrac{\partial V_{a\alpha}^\text{Cu-H}(r_{a\alpha})}{\partial r_{a\alpha}}.\dfrac{\partial r_{a\alpha}}{\partial X_{\alpha 
i}}&=&\Bigg[ \frac{1}{2} \sin
 \left(\frac{\pi(r_{a\alpha}-b_2)}{b_2-b_1}\right)  \left(\frac{\pi}{b_2-b_1} \right) \left(V_\text{Ryd}(r_{a\alpha}) -V_\text{Ryd}(b_2)
  \right) + (1-\rho(r_{a\alpha}))\nonumber\\
  &&\Big[-lV_\text{Ryd}(r_{a\alpha})-\exp
\lbrace-l(r_{a\alpha}-z)\rbrace\left[c_{1} + 2c_{2}(r_{a\alpha}-z)+3c_{3}(r_{a\alpha}-z)^{2}\right]\Big]\Bigg]\nonumber \\ 
&&\times \left(\frac{X_{\alpha i}-X_{ai}}
{r_{a\alpha}}\right)
\label{1st2nd}
\end{eqnarray}

\subsubsection{If $r_{a\alpha}$ $>$ $b_{2}$:}

In this range, $\rho(r_{a\alpha})$ = 1 \; and \; $\frac{\partial \rho (r_{a\alpha})}{\partial r_{a\alpha}}$ = 0.
\begin{eqnarray}
\frac{\partial V^\text{Cu-H}_{a\alpha}(r_{a\alpha})}{\partial r_{a \alpha}} = \dfrac{\partial V_\text{Ryd}(b_2)}{\partial r_{a\alpha}}= 0,
\end{eqnarray}
since $V_\text{Ryd}(b_2)$ is independent of the gas-metal atom distance $r_{a\alpha}$.
\begin{eqnarray}
\therefore \dfrac{\partial V_{a\alpha}^\text{Cu-H}(r_{a\alpha})}{\partial r_{a\alpha}}.\dfrac{\partial r_{a\alpha}}{\partial X_{\alpha i}}=  0
\label{1st3nd}
\end{eqnarray}

\clearpage
\bibliography{article}
	
\end{document}